\documentstyle[12pt]{article}
\begin{document}

\newcommand \s[1] {\mbox{\hspace{#1cm}}}
\newcommand \FR[2]
 {\s{.06}{\strut\displaystyle#1\over\strut\displaystyle#2}\s{.06}}
\newcommand \fr[2] {\s{.06}{\textstyle{#1\over#2}}\s{.06}}
\newcommand \bin[2] {\left( \begin{array}c #1\\ #2 \end{array} \right) }
\newcommand \eqnum[1]{\eqno (#1)}
\renewcommand \d {{\rm d}}
\newcommand \e {{\rm e}}
\renewcommand \i {{\rm i}}
\newcommand \Mu {{\sl M}}
\newcommand \Nu {{\sl N}}

\hyphenation {sca-lar spin-or coun-ter-term}

\setcounter{page}0
\thispagestyle{empty}
\begin{center} {\Large\bf Difficulties of an infrared extension
 \vspace{5mm} \\
 of differential renormalization} \\
 \vfill

 {\large L.V.Avdeev\footnote{E-mail: avdeevL@theor.jinrc.dubna.su},~
  D.I.Kazakov\footnote{E-mail: kazakovD@theor.jinrc.dubna.su}~ and~
  I.N.Kondrashuk\footnote{E-mail: ikond@theor.jinrc.dubna.su}} \\
 \vspace{1cm}
 {\em Laboratory of Theoretical Physics \\
  Joint Institute for Nuclear Research \\
  141980 Dubna (Moscow Region)\\
  Russian Federation} \\
\end{center}
\vfill

\begin{abstract}
 We investigate the possibility of generalizing differential
 renormalization of D.Z.Freedman, K.Johnson and J.I.Latorre in an
 invariant fashion to theories with infrared divergencies via an
 infrared $\tilde{R}$ operation. Two-dimensional $\sigma$ models and the
 four-dimensional $\phi^4$ theory diagrams with exceptional momenta are
 used as examples, while dimensional renormalization serves as a test
 scheme for comparison. We write the basic differential identities of
 the method simultaneously in co-ordinate and momentum space,
 introducing two scales which remove ultraviolet and infrared
 singularities. The consistent set of Fourier-transformation formulae is
 derived. However, the values for tadpole-type Feynman integrals in
 higher orders of perturbation theory prove to be ambiguous, depending
 on the order of evaluation of the subgraphs. In two dimensions, even
 earlier than this ambiguity manifests itself, renormalization-group
 calculations based on infrared extension of differential
 renormalization lead to incorrect results. We conclude that the
 extended differential renormalization procedure does not perform the
 infrared $\tilde{R}$ operation in a self-consistent way, as the
 original recipe does the ultraviolet $R$ operation.
\end{abstract}
\vfill

\pagebreak

\section{Introduction} \indent

Differential renormalization \cite{Dif} can successfully perform an
ultraviolet $R$ operation by means of replacing essentially singular
quantities in co-ordinate space with derivatives of less singular
expressions, and after that, by integrating the derivatives by parts in
order to turn them into powers of external momenta. Then, at first
sight, one sees no serious obstacles --- it would just seem very natural
indeed --- that the same trick with singular expressions in momentum
space could likewise automatically perform an infrared $\tilde{R}$
operation \cite{R*}.

A need of the latter in renormalization-group calculations (leaving
alone operator-product and other asymptotic expansions) may arise of two
causes. In the first place, a quantum field theory may be {\em
intrinsically} infrared-dangerous owing to the fact that the theory with
zero masses may not strictly exist at all as, for example, a massless
scalar field in two dimensions. However, real computations with nonzero
masses are always exceedingly cumbersome technically while, on the
other hand, we know that in the so-called ``massless'' renormalization
schemes \cite{Mless} (like the minimal subtraction scheme
\cite{MS,Speer} of dimensional or analytic regularization) there is no
essential dependence of renormalization-group functions on dimensional
parameters (except for a possible mixing in renormalizations of the
masses themselves). Thus, we would like to set them all to zero from the
very beginning but use an infrared $\tilde{R}$ operation to avoid the
false infrared singularities.

In the second case, {\em artificial} infrared divergencies may be caused
by a singular choice of external momenta of a Green's function, quite
well defined in a general situation. For technical reasons, such a
choice may seem more convenient, or it may even be the only available
choice that would allow us to compute a complicated multiloop
vertex-type graph analytically by means of an infrared rearrangement
\cite{IR-re}.

Although there are no theoretical grounds, like the causality principle
of the local quantum field theory, for attributing a physical meaning to
the infrared $\tilde{R}$ operation in itself --- the reasons look quite
technical both with intrinsic and artificial infrared divergencies ---
however, in many practically important cases its use may be crucial for
achieving computational results \cite{Multi,sig5}.

Introducing the infrared $\tilde{R}$ operation into the formalism of
differential renormalization, we should keep track of three important
points. First of all, the scheme should retain its {\em
self-consistency:} the results should never depend on the order of
manipulations we perform. The second requirement is that we should
distinguish {\em two dimensional scales:} a new infrared renormalization
scale $\nu$, and an ultraviolet scale $\mu$ already present in the
formalism. Otherwise, infrared logarithms would be the same as
ultraviolet logarithms, and we would never be able to extract from the
finite renormalized Feynman diagrams the correct information about
renormalization-group functions, contained in their $\mu$ dependence.
And last but by no means least point is the {\em invariance} of the
regularization \cite{Inv}, which ensures the fact that formal symmetries
of a quantum field theory are retained after renormalization
\cite{Speer,Action}.

We formulate a natural generalization of differential renormalization to
theories where, besides usual ultraviolet singularities, intrinsic or
artificial infrared divergencies do appear. Our particular attention is
paid to two-dimensional $\sigma$ models and to the four-dimensional
$\phi^4$ theory. Throughout our renormalization-group calculations, we
compare the results of the extended differential renormalization with
the minimal subtraction scheme of dimensional regularization as an
example of an undoubtedly valid, though different, renormalization
scheme in the range of scheme arbitrariness.

\section{D=2: Intrinsic infrared divergencies} \indent

As the first example, we consider two-dimensional $\sigma$ models
\cite{sig1,sig2} where we meet infrared difficulties of both
intrinsic and artificial origin.

The starting point of differential renormalization in a two-dimensional
theory is the following replacement which effectively subtracts an
infinite local ultraviolet counterterm proportional to the $\delta$
function in co-ordinate space:
\begin{equation} 1/x^2 = \fr 1 8 \Box_x~ \ln^2 (x^2 \Mu^2)~, \label{x2}
\end{equation}
where $\Box_x$ denotes $(\partial/\partial x_\mu)^2$ (we work in
Euclidean space). The other way of looking at the matter is to consider
eq.(\ref{x2}) as a real equality under a regularization, bearing in mind
that the regularized expressions are renormalized afterwards by minimal
subtractions which simply turn the regulator $\Mu$$\to$$\infty$ into a
finite renormalization scale.

With the aid of eq.(\ref{x2}), integrating the derivatives by parts and
abandoning surface terms (which is one more prescription of the
regularization), we can derive a basic set of formulae for Fourier
transformations. We start with the ultraviolet-divergent integrals of
the form
\begin{equation} I_n = \int \d^2 x~ \e^{\i p x}~
 \FR {\ln^n (x^2 \Mu^2)} {x^2}. \label{I}
\end{equation}
Replacing $1/x^2$ through eq.(\ref{x2}), we integrate one $\partial /
\partial x_\mu$ by parts. Of course, we suppose the Leibniz
differentiation rule to remain valid under the regularization. The first
derivative of the logarithm is a nonsingular function in two dimensions,
and hence, it can be substituted explicitly as $2x_\mu/x^2$ without any
possible $\delta$-function-like terms. The resulting equation can be
solved with respect to $I_n$:
$$ I_n = - \FR {\i p_\mu} {2(n+1)}
 \int \d^2 x~ \e^{\i p x}~ \FR {x_\mu} {x^2} \ln^{n+1} (x^2 \Mu^2)~. $$
The integral on the right-hand side has no dangerous singularities, and
we can unambiguously evaluate it, using an intermediate analytic
regularization which is taken off in the end,
\begin{eqnarray} \int &\d^2 x& \e^{\i p x}~ \FR {x_\mu} {x^2}
 \ln^n (x^2 \Mu^2) \label{x1}\\*
&=&
 \left[ (-)^n \FR {\d^n} {\d\alpha^n}
  \FR 1 \i \FR \partial {\partial p_\mu}
  \FR 1 {\left( \Mu^2 \right)^\alpha}
  \int \d^2 x \FR {\e^{\i p x}} {\left( x^2 \right)^{1+\alpha}}
 \right] _{\displaystyle \alpha=0} \nonumber\\
&=& 2 \pi \i~ \FR {p_\mu} {p^2} (-)^n
 \left. \FR {\d^n} {\d\alpha^n}
  \left[ \left( p^2/\mu^2 \right)^\alpha F(\alpha)
  \right]
 \right| _{\displaystyle \alpha=0} \nonumber\\
&=& 2 \pi \i~ \FR {p_\mu} {p^2} (-)^n \sum_{m=0}^n
 \bin n m F_{n-m}~ \ln^m (p^2/\mu^2)~, \nonumber \\[4mm]
I_n &=& \pi \FR {(-)^{n+1}} {n+1} \sum_{m=0}^{n+1}
 \bin{n+1}m F_{n+1-m}~ \ln^m (p^2/\mu^2)~, \label{I=}
\end{eqnarray}
where $\mu^2=4~\Mu^2~\e^{-2C}$,
\begin{equation} F(\alpha) = \e^{-2C\alpha}~
 \FR {\Gamma(1-\alpha)} {\Gamma(1+\alpha)} = \exp
 \left[ 2 \sum_{m=1}^\infty \FR {\zeta(2m+1)} {2m+1} \alpha^{2m+1}
 \right] , \label{Falpha}
\end{equation}
$F_m$ are $m$th derivatives of that function at zero,
\begin{eqnarray*} &F_0=1, \s1 F_1=F_2=0, \s1 F_3=4~\zeta(3), \s1
 F_4=0, \s1 F_5=48~\zeta(5),& \\*
&F_6=160~\zeta^2 (3), \s1 F_7=1440~\zeta(7), \s1 \ldots ~ ,&
\end{eqnarray*}
$\zeta(m)$ is the Riemann zeta function, and $C$ is the Euler
constant.

Since the initial integrals $I_n$, eq.(\ref{I}), were logarithmically
divergent, in the most general situation we could add to the
renormalized expressions (\ref{I=}) some constants $C_n$, reflecting a
local arbitrariness. In the context of differential renormalization, we
could leave an explicit $\delta$-function term in eq.(\ref{x2}). That
would simply be equivalent to redefining the renormalization scale
$\Mu$. Therefore, we have chosen to absorb the arbitrariness into the
definition of $\Mu$, and no more $C_n$ could ever appear with the
minimal subtractions on eq.(\ref{I=}).

Now, we can try to execute precisely the same procedure for infrared
rather than ultraviolet divergencies, starting from momentum space with
an infrared regulator mass $\nu$$\to$0, which becomes another finite
renormalization scale after the minimal subtractions:
\begin{equation} 1/p^2 = \fr 1 8 \Box_p~ \ln^2 (p^2/\nu^2)~,
 \label{p2}
\end{equation}
\begin{equation} \int \d^2 p~ \e^{-\i p x}~ \FR {p_\mu} {p^2}
 \ln^n (p^2/\nu^2) = -2 \pi \i~ \FR {x_\mu} {x^2} (-)^n \sum_{m=0}^n
 \bin n m F_{n-m}~ \ln^m (x^2 \Nu^2)~, \label{p1}
\end{equation}
\begin{equation}
 i_n = \int \d^2 p~ \e^{-\i p x}~ \FR {\ln^n (p^2/\nu^2)} {p^2}
 = \pi \FR {(-)^{n+1}} {n+1} \sum_{m=0}^{n+1} \bin{n+1}m F_{n+1-m}~
 \ln^m (x^2 \Nu^2)~. \label{i=}
\end{equation}
Again, we have denoted $\Nu^2=\fr{1}4\nu^2~\e^{2C}$. (Everywhere,
$\Mu^2$ and $\Nu^2$ accompany $x^2$, while $p^2$ is divided by $\mu^2$
or $\nu^2$).

\subsection{The simplest example} \indent

The fact that there really is an infrared difficulty for differential
renormalization in two dimensions is quite obvious from considering the
simplest one-loop tadpole diagram which appears in $\sigma$ models
(analyzed below in subsects.\ref{susy} and \ref{ren}):
\begin{equation}
 \begin{picture}(10,8.5)(-5,-2.5) 
  \put(0,0){\circle 7}
  \put(0,-3.5){\line(-1,-1){3.6}}
  \put(0,-3.5){\line(1,-1){3.6}}
  \put(0,-3.5){\circle*1}
 \end{picture}
 = \int \d^2 p \FR 1 {p^2} . \label{ex1}
\end{equation}
The original method of ref.\cite{Dif} would simply set this diagram to
zero, leaving no trace of any ultraviolet logarithm. Therefore, in order
to find a correct contribution to the $\beta$ function, we need a
special procedure for defining such integrals which involve no external
momenta or co-ordinates and are both ultraviolet- and
infrared-divergent.

Let us start with the outline of our ideas in application to this
integral as a simple example. We always assume that the whole scheme can
be treated as an invariant regularization procedure which allows formal
manipulations with divergent integrals. Then, we can identically
transform eq.(\ref{ex1}) to introduce an auxiliary external momentum
argument $k_\mu$:
\begin{equation} \int \d^2 p \FR 1 {p^2} = k_\mu \int \d^2 p \FR 1 {p^2}
 \FR {\left(k-p\right)_\mu} {\left(k-p\right)^2}
 - \int \d^2 p \FR {p_\mu} {p^2}
 \FR {\left(k-p\right)_\mu} {\left(k-p\right)^2} . \label{ex2}
\end{equation}
The first integral on the right-hand side of eq.(\ref{ex2}) is
infrared-divergent but ultraviolet-regular, while the second integral
has only an ultraviolet singularity. Now we use the Fourier
transformation to rewrite eq.(\ref{ex2}) as
\begin{equation} \int \FR {\d^2 x} {\left(2\pi\right)^2} \e^{\i k x}
 \Bigl( k_\mu \int \d^2 p \FR {\e^{-\i p x}} {p^2}
  - \int \d^2 p~ \e^{-\i p x} \FR {p_\mu} {p^2}
 \Bigr) \int \d^2 q~ \e^{-\i q x} \FR {q_\mu} {q^2} . \label{ex3}
\end{equation}
One Fourier integral is regular [it is the simplest particular case of
eq.(\ref{p1})]
\begin{equation} \int \d^2 p~ \e^{-\i p x} \FR {p_\mu} {p^2}
 = - 2\pi\i \FR {x_\mu} {x^2} . \label{ex4}
\end{equation}
The other Fourier integral in eq.(\ref{ex3}) requires an infrared
renormalization. It is $i_0$, eq.(\ref{i=}),
\begin{equation} \int \d^2 p \FR {\e^{-\i p x}} {p^2}
 = - \pi~ \ln (x^2 \Nu^2)~ . \label{ex5}
\end{equation}
In the extended differential renormalization, the latter equation is
derived from the momentum-space differential identity (\ref{p2}) via
integration by parts and abandoning surface terms which are absorbed
into local $p$-space counterterms in the formalism of the infrared
$\tilde{R}$ operation \cite{R*}. The result involves a new infrared
renormalization scale. For the case of the Fourier transform of a free
scalar-field propagator (\ref{ex5}), the physical meaning of the
$\tilde{R}$ operation can easily be understood: it recovers the most
essential (logarithmic) part of the massive propagator that remains as
the mass goes to zero
$$ \int \d^2 p \FR {\e^{-\i p x}} {p^2+m^2} = 2\pi~K_0\left(m|x|\right)
 = - \pi~ \ln \left( \fr 1 4 m^2 x^2 \right) + o(1)~, \s1 m\to 0~ . $$
In this interpretation, the infrared differential regularization
parameter in eq.(\ref{p2}) is proportional to the mass
$\nu^2$=$m^2\e^{-2C}$. However, the general $\tilde{R}$ formalism is not
reduced to introducing a mass. It renormalizes any infrared-divergent
diagram in a dual way to the ordinary $R$ operation with local $x$-space
counterterms, the structure being completely analogous, but with the
$x$$\leftrightarrow$$p$ interchange.

To finish the computation of the tadpole diagram (\ref{ex1}), we
substitute eqs.(\ref{ex4}) and (\ref{ex5}) into eq.(\ref{ex3}). The
integrals over $x$ are taken by the original rules of differential
renormalization, which lead to eqs.(\ref{x1}) and (\ref{I=}), in
particular,
$$ \int \d^2 x~ \e^{\i k x} \FR {x_\mu} {x^2} \ln (x^2 \Nu^2)
 = - 2\pi\i \FR {k_\mu} {k^2} \ln \FR {k^2} {\nu^2} , \s1
 \int \d^2 x \FR {\e^{\i k x}} {x^2} = - \pi~ \ln \FR {k^2} {\mu^2} , $$
where the ultraviolet scale $\mu$ has appeared. In the resulting
expression for the diagram (\ref{ex1}) the dependence on the auxiliary
momentum cancels, and we obtain $\pi~\ln(\mu^2/\nu^2)$. Differentiating
this with respect to $\ln\mu^2$, we reproduce the correct contribution
to the $\beta$ function.

A systematic study of the implications of the infrared differential
renormalization is presented below. First, in subsect.\ref{fou}, we try
to derive a maximum set of relations (including singular functions) from
the requirements of invariance of regularization. Since the maximum set
entails a contradiction (subsect.\ref{amb}), we restrict ourselves to
avoid any explicit singular differetiation. However, even without that,
we recover all the essential consequences, so that the more cautious
approach turns to be equivalent to the unrestricted one.

\subsection{The consistent set of Fourier integrals} \label{fou} \indent

Having added the momentum-space differential identity (\ref{p2}), we
ought to check the consistency of the scheme as a whole. We expect of a
good regularization that it is invariant \cite{Inv}, that is, permits us
to do with our integrals some formal manipulations, which is important
for retaining symmetries of quantum field models. One of those features
is the possibility of differentiating the integrals with respect to
external parameters, for example, the Fourier-transformed argument, to
compute other integrals, with external Lorentz indices, as we have done
in deriving eq.(\ref{x1}). Other natural requirements are the
possibilities of changing the order of integrations, of shifting
integration variables, and of identically transforming subintegral
expressions, like expanding $(p$$-$$q)^2$ or canceling the numerator
with the denominator. Such transformations are often performed, in
particular, when one does superfield algebra in supersymmetric models to
reduce supergraphs to usual Feynman integrals.

Believing that the extended differential regularization can be treated
as an invariant overall regularization, let us apply $-$$\Box_x$ to
$i_n$ to obtain, after canceling the denominator $p^2$ in eq.(\ref{i=}),
new integrals
\begin{equation} j_n = -\Box_x~ i_n
 = \int \d^2 p~ \e^{-\i p x}~ \ln^n (p^2/\nu^2)~, \label{j}
\end{equation}
which contain no infrared divergencies, as differs from eq.(\ref{i=}).
Note also that any arbitrary renormalization-specific constants $c_n$
added to $i_n$ are annihilated by differentiation, so that $j_n$ will in
any case be determined uniquely.

At $n$=0, we must evidently get $j_0$=$4\pi^2\delta_2(x)$. On the other
hand, differentiating the right-hand side of eq.(\ref{i=}) for $i_0$,
produces the laplacian of the logarithm. Hence,
\begin{equation} \Box_x~ \ln (x^2 \Nu^2) = 4 \pi~ \delta_2 (x)~.
 \label{d2}
\end{equation}
Of course, in fact, eq.(\ref{d2}) is true without any reference to
differential regularization ($\Nu$ drops out, too). The identity can as
well be derived by some other means, for example, by explicitly
introducing the causal rule of passing over $x^2$=0, expanding the
left-hand side of eq.(\ref{d2}), and directly evaluating the integral of
it. The extended differential regularization reproduces the result
correctly.

Now we proceed to $n$=1. The laplacian applied to the logarithm squared
in $i_1$, eq.(\ref{i=}), gives us two terms: when both derivatives act
on one logarithm, we use eq.(\ref{d2}), while the nonsingular first
derivatives of the logarithm can be substituted directly. Thus, we get
\begin{equation} j_1 = -\Box_x~ i_1
 ~=~ - 4 \pi^2~ \delta_2 (x)~ \ln (x^2 \Nu^2) - 4 \pi/x^2. \label{j1}
\end{equation}
To attach a meaning to eq.(\ref{j1}), we need to extend the definition
of the logarithm to zero point in co-ordinate space. However, our set of
formulae already includes such a definition: if we expand, on the
right-hand side of eq.(\ref{x2}), the action of the laplacian, we get
besides $1/x^2$ an additional term which equals $\pi\delta_2(x) \ln(x^2
\Mu^2)$ in view of eq.(\ref{d2}). We have decided that by definition of
$\Mu$ this local term is absent in eq.(\ref{x2}); therefore,
\begin{equation} \delta_2 (x)~ \ln x^2 = \delta_2 (x)~ \ln (1/\Mu^2)~.
 \label{x0}
\end{equation}
It is this formula that shows us explicitly what the regularization
really changes. For the rest, everything looks as if nothing has been
regularized at all, provided we manage to take proper care of singular
functions.

As soon as we have established eqs.(\ref{x0}) and (\ref{d2}), we can
uniformly obtain all the integrals $j_n$, eq.(\ref{j}), from $i_n$,
eq.(\ref{i=}), arriving at the general formula

\pagebreak[1]
$$ j_n = \int \d^2 p~ \e^{-\i p x}~ \ln^n (p^2/\nu^2) ~=~ 4 \pi~ (-)^n
 \left[ \pi \delta_2 (x) \sum_{m=0}^n \bin n m F_{n-m}~
  \ln^m (\Nu^2/\Mu^2)
 \right. $$
\begin{equation}
 \left. +~ \FR n {x^2} \sum_{m=0}^{n-1} \bin{n-1}m F_{n-1-m}~
  \ln^m (x^2 \Nu^2)
 \right] . \label{j=}
\end{equation}

Quite analogously, acting with $-$$\Box_p$ on eq.(\ref{I=}) and
remembering eq.(\ref{p2}), we deduce the $x$$\leftrightarrow$$p$
complementary formulae
$$ \Box_p~ \ln (p^2/\mu^2) = 4 \pi~ \delta_2 (p)~, \eqnum{\ref{d2}'} $$
\begin{equation} \delta_2 (p)~ \ln p^2 = \delta_2 (p)~ \ln \nu^2,
 \label{p0}
\end{equation}
\pagebreak[1]
$$ J_n = \int \d^2 x~ \e^{\i p x}~ \ln^n (x^2 \Mu^2) ~=~ 4 \pi~ (-)^n
 \left[ \pi \delta_2 (p) \sum_{m=0}^n \bin n m F_{n-m}~
  \ln^m (\nu^2/\mu^2)
 \right. $$
\begin{equation}
 \left. +~ \FR n {p^2} \sum_{m=0}^{n-1} \bin{n-1}m F_{n-1-m}~
  \ln^m (p^2/\mu^2)
 \right] . \label{J=}
\end{equation}
Evidently, the ratios $\Nu^2/\Mu^2$ in eq.(\ref{j=}) and $\nu^2/\mu^2$
in eq.(\ref{J=}) are the same.

Another observation is that, as $J_n$ is ultraviolet regular, it does
not generate an ultraviolet scale parameter by itself. Therefore, $\Mu$
in eq.(\ref{J=}) is just a dummy argument (strictly related to $\mu$ on
the right-hand side) while the real stamp of the regularization on
$J_n$ is the infrared scale $\nu$ from eq.(\ref{p0}). The
self-consistency of replacing $\Mu$ in eq.(\ref{J=}) can be checked
directly by re-expanding the logarithm with an arbitrary mass argument
$\Lambda$ by the Newton's binomial, and resumming the right-hand side.
The result

\pagebreak[1]
$$ \int \d^2 x~ \e^{\i p x}~ \ln^n (x^2 \Lambda^2) ~=~ 4 \pi~ (-)^n
 \left[ \pi \delta_2 (p) \sum_{m=0}^n \bin n m F_{n-m}~
  \ln^m (\nu^2/\lambda^2)
 \right. $$
\begin{equation}
 \left. +~ \FR n {p^2} \sum_{m=0}^{n-1} \bin{n-1}m F_{n-1-m}~
  \ln^m (p^2/\lambda^2)
 \right] \label{JL}
\end{equation}
actually looks the same as eq.(\ref{J=}). Likewise, we can freely
change $\nu$ in eq.(\ref{j=}) to a dummy variable $\lambda$ which can
even be set equal to $\mu$,

\pagebreak[1]
$$ \int \d^2 p~ \e^{-\i p x}~ \ln^n (p^2/\lambda^2) ~=~ 4 \pi~ (-)^n
 \left[ \pi \delta_2 (x) \sum_{m=0}^n \bin n m F_{n-m}~
  \ln^m (\Lambda^2/\Mu^2)
 \right. $$
\begin{equation}
 \left. +~ \FR n {x^2} \sum_{m=0}^{n-1} \bin{n-1}m F_{n-1-m}~
  \ln^m (x^2 \Lambda^2)
 \right] . \label{jl}
\end{equation}
The intermediate integrals with Lorentz indices, eqs.(\ref{x1}) and
(\ref{p1}), are both ultraviolet- and infrared-regular. Thus, they
introduce no scales at all, $\mu$ and $\nu$ being their dummy arguments.

The next step in checking the consistency is to take the Fourier
transforms of eqs.(\ref{JL}) and (\ref{jl}). One easily sees that the
resulting formulae allow us to recursively recover the very first
integrals: $I_n$, eq.(\ref{I}), and $i_n$, eq.(\ref{i=}). The double
circle closes. By induction, using binomial re-expansions, we can prove
that formulae (\ref{I=}) and (\ref{i=}) are exactly reproduced
without any $C_n$ and $c_n$, which reflects the fact that we made
eqs.(\ref{x0}) and (\ref{p0}) agree with the differential relations
(\ref{x2}) and (\ref{p2}). For other masses under the logarithms, we get
quite definite additive terms,
\begin{eqnarray} \int &\d^2 x&
 \e^{\i p x}~ \FR {\ln^n (x^2 \Lambda^2)} {x^2}~ \label{IL} \\*
&=& \FR \pi {n+1}
 \left[ (-)^{n+1} \sum_{m=0}^{n+1} \bin{n+1}m F_{n+1-m}~
  \ln^m (p^2/\lambda^2) - \ln^{n+1} (\lambda^2/\mu^2)
 \right] , \nonumber \\[4mm]
\int &\d^2 p&
 \e^{-\i p x}~ \FR {\ln^n (p^2/\lambda^2)} {p^2} \label{il} \\*
&=& \FR \pi {n+1}
 \left[ (-)^{n+1} \sum_{m=0}^{n+1} \bin{n+1}m F_{n+1-m}
  \ln^m (x^2 \Lambda^2) - \ln^{n+1} (\nu^2/\lambda^2)
 \right] , \nonumber
\end{eqnarray}
with $\lambda^2=4~\Lambda^2~\e^{-2C}$ in eqs.(\ref{JL})--(\ref{il}).

\subsection{Ambiguities in repeated integrals} \label{amb} \indent

Everything looks fine at the moment, and we could start to compute
diagrams. However, the point is that we actually need a little more than
just switching to the Fourier transforms and back. Namely, we expect
that a property similar to associativity of convolution will hold.
Suppose, there are three quantities
\begin{equation} a_n (p) = \int \d^2 x~ \e^{\i p x}~ A_n (x)~, \s1
 A_n (x) = \int \FR {d^2 p} {(2\pi)^2} \e^{-\i p x}~ a_n (p)~,
 \label{an}
\end{equation}
($n$=1,2,3), and we are going to evaluate the integral of their product
in momentum space. We can represent it in three different ways,
substituting eq.(\ref{an}) either for $a_1$, $a_2$, or $a_3$:

\begin{eqnarray} X &=& \int \d^2 p~ a_1(p)~ a_2(p)~ a_3(p) \nonumber\\*
&=& \int \d^2 x~ A_1(x)
 \int \d^2 p~ \e^{-\i p x}~ a_2(p)~ a_3(p) \label{A1} \\*
&=& \int \d^2 x~ A_2(x)
 \int \d^2 p~ \e^{-\i p x}~ a_1(p)~ a_3(p) \label{A2} \\*
&=& \int \d^2 x~ A_3(x)
 \int \d^2 p~ \e^{-\i p x}~ a_1(p)~ a_2(p)~, \label{A3}
\end{eqnarray}
where the integral over $x$ can be evaluated by a trick of inserting a
unit
\begin{equation} \int \d^2 x~ \ldots~ = \int \d^2 q~ \delta_2(q)
 \int \d^2 x~ \e^{\i q x}~ \ldots \label{trick}
\end{equation}
and subsequently applying eq.(\ref{p0}). Either way
(\ref{A1})--(\ref{A3}) should lead to the same result in the framework
of a consistent regularization. Let us, however, substitute particular
expressions
\begin{equation} a_1(p) = \FR {2~\ln(p^2/\nu^2)} {p^2} , \s1
 a_2=a_3=\ln(p^2/\mu^2)~. \label{choice}
\end{equation}
All the integrals that arise can be evaluated with the aid of
eqs.(\ref{JL})--(\ref{il}), after products of different logarithms are
re-expanded in powers of one of them. The results are
\begin{eqnarray} X &\stackrel {\strut (\ref{A1})} \Rightarrow& \pi
 \Bigl[ ~\fr 1 6 \ln^4 (\nu^2/\mu^2)
  + \fr 8 3 \zeta(3)~ \ln (\nu^2/\mu^2)~
 \Bigr] , \label{X1} \\*
X &\stackrel {\strut (\ref{A2}),(\ref{A3})} \Longrightarrow& \pi
 \Bigl[ ~\fr 1 6 \ln^4 (\nu^2/\mu^2)
  - \fr 4 3 \zeta(3)~ \ln (\nu^2/\mu^2)~
 \Bigr] . \label{X2}
\end{eqnarray}
We see that there are two ways of computing the integral, which give us
two different answers. The resulting expressions (\ref{X1}) and
(\ref{X2}) coincide only if we set $\nu^2$=$\mu^2$, that is, use the
same scale for renormalizing both ultraviolet and infrared divergencies.
As has been said at the beginning, this would not be satisfactory for
renormalization-group applications.

One may suspect that the inconsistency has resulted from undue liberty
in our dealing with singular functions. Indeed, in establishing our set
of formulae, we exceeded the necessary minimum of operations, essential
for evaluating Feynman integrals. So, in principle, we can do without
explicit differentiation the result of which might involve a singular
function.

Then, we ought to exclude all the singular relations, (\ref{d2}),
(\ref{j1}), (\ref{x0}), and (\ref{p0}), thus, being unable to get $j_n$,
eq.(\ref{j}), from $i_n$. Nevertheless, the basic divergent integrals
themselves $I_n$ and $i_n$, eqs.(\ref{I=}) and (\ref{i=}), can be
derived from the differential identities eqs.(\ref{x2}) and (\ref{p2})
through intermediate finite integrals, eqs.(\ref{p1}) and (\ref{x1}), as
before. Re-expanding the logarithms, we reproduce the integrals with
arbitrary mass arguments, eqs.(\ref{IL}) and (\ref{il}); and finally,
eqs.(\ref{jl}) and (\ref{JL}) are obtained as their resummed Fourier
transforms with the particular cases of eqs.(\ref{j=}) and (\ref{J=}).

Although the trick of eq.(\ref{trick}) becomes of no use, we can, all
the same, define the general external-parameter-independent integrals in
eqs.(\ref{A1})--(\ref{A3}) by utilizing the assumed invariance of the
regularization:

\begin{eqnarray} \int &\d^2 x& \FR {\ln^n (x^2\Lambda^2)} {x^2}
 = \int \d^2 x \FR {(y-x)\cdot y - (y-x)\cdot x} {(y-x)^2~~ x^2}
 \ln^n (x^2\Lambda^2) \label{tx} \\*
&=& \int \FR {\d^2 p} {(2\pi)^2} \e^{-\i p y}
 \left( \int \d^2 z~ \e^{\i p z}~ \FR {z_\mu} {z^2}
 \right)
 \left[ \int \d^2 x~ \e^{\i p x}
  \left( \FR {y_\mu} {x^2} - \FR {x_\mu} {x^2}
  \right) \ln^n (x^2\Lambda^2)
 \right] \nonumber\\*
&=& \FR \pi {n+1}
 \left[ (-)^{n+1} \sum_{m=0}^{n+1} \bin {n+1} m F_{n+1-m}~
  \ln^m (\Nu^2/\Lambda^2) - \ln^{n+1} (\Lambda^2/\Mu^2)
 \right] , \nonumber
\end{eqnarray}
where we have made use of eqs.(\ref{x1}), (\ref{IL}) and (\ref{il}).
Note that the term with $m$=0 is always present, even at
$\Lambda^2$=$\Nu^2$. In the same way, we can evaluate the corresponding
momentum integrals

\begin{eqnarray} \int &\d^2 p& \FR {\ln^n (p^2/\lambda^2)} {p^2}
 \label{tp}\\*
&=& \FR \pi {n+1}
 \left[ (-)^{n+1} \sum_{m=0}^{n+1} \bin {n+1} m F_{n+1-m}~
  \ln^m (\lambda^2/\mu^2) - \ln^{n+1} (\nu^2/\lambda^2)
 \right] . \nonumber
\end{eqnarray}

Now, for the chosen $a_1$, $a_2$, and $a_3$, as in eq.(\ref{choice}),
the integrals on the right-hand sides of eqs.(\ref{A1})--(\ref{A3}) are
directly calculable, and we reproduce the two different results
(\ref{X1}) and (\ref{X2}) again.

\subsection{N=2 supersymmetric sigma model} \label{susy} \indent

\begin{figure}[htbp] \centering
 \begin{picture}(24,23)(-10,-13) 
  \put(0,0){\circle{14}}
  \put(0,7){\line(-1,-2){5.6}} 
  \put(0,7){\line(1,-2){5.6}}
  \put(-5.6,-4.2){\line(1,0){11.2}}
  \put(-5.6,4.2){\vector(1,2){.3}} 
  \put(-2.8,1.4){\vector(1,2){.3}}
  \put(-4.2,2.8){\circle*{.6}}
  \put(0,-4.2){\vector(1,0){.7}} 
  \put(0,-7){\vector(1,0){.7}}
  \put(0,-5.6){\circle*{.6}}
  \put(0,7){\circle*1} 
  \put(-5.6,-4.2){\circle*1}
  \put(5.6,-4.2){\circle*1}
  \put(0,7){\line(0,1)3} 
  \put(-5.6,-4.2){\line(-2,-1){3.6}}
  \put(5.6,-4.2){\line(2,-1){3.6}}
 \end{picture}
 (a)\s3
 \begin{picture}(20,20)(-10,-13) 
  \put(3.6,3.6){\oval(5.9,5.9)[t]}
  \put(3.6,3.6){\oval(5.9,5.9)[rb]}
  \put(-3.6,3.6){\oval(5.9,5.9)[t]}
  \put(-3.6,3.6){\oval(5.9,5.9)[lb]}
  \put(-3.6,-3.6){\oval(5.9,5.9)[b]}
  \put(-3.6,-3.6){\oval(5.9,5.9)[lt]}
  \put(3.6,-3.6){\oval(5.9,5.9)[b]}
  \put(3.6,-3.6){\oval(5.9,5.9)[rt]}
  \put(-.6,-3.6){\line(1,6){1.2}}
  \put(.6,-3.6){\line(-1,6){1.2}}
  \put(-3.6,-.6){\line(6,1){7.2}}
  \put(-3.6,.6){\line(6,-1){7.2}}
  \put(0,0){\circle*1}
 \end{picture}
 (b)
 \caption{The divergent four-loop momentum integrals in the $N$=2
  supersymmetric two-dimensional $\sigma$ model. Lines represent $1/p^2$
  propagators, arrows stand for additional $p_\mu$ in the numerator, and
  small dots denote contractions of Lorentz indices.}

 \label{4loop}
\end{figure}
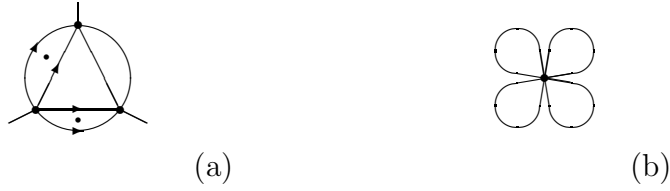

Yet, let us see how grave this contradiction is from the practical
viewpoint of computing Feynman diagrams. We consider first the four-loop
approximation in the $N$=2 supersymmetric two-dimensional $\sigma$
model. After the supergraph algebra in the background-field formalism is
done \cite{sig2}, there appears just one nontrivial divergent four-loop
momentum integral which is presented in fig.\ref{4loop}a. The rest of
the divergent integrals are of the primitive tadpole type
(fig.\ref{4loop}b); we leave them alone for the moment.

\begin{figure}[htbp] $$
 \begin{picture}(24,11)(-11,-1) 
  \put(0,0){\circle{14}}
  \put(0,7){\line(-1,-2){5.6}} 
  \put(0,7){\line(1,-2){5.6}}
  \put(-5.6,-4.2){\line(1,0){11.2}}
  \put(-5.6,4.2){\vector(1,2){.3}} 
  \put(-2.8,1.4){\vector(1,2){.3}}
  \put(-4.2,2.8){\circle*{.6}}
  \put(0,-4.2){\vector(1,0){.7}} 
  \put(0,-7){\vector(1,0){.7}}
  \put(0,-5.6){\circle*{.6}}
  \put(0,7){\circle*1} 
  \put(-5.6,-4.2){\circle*1}
  \put(5.6,-4.2){\circle*1}
  \put(0,7){\line(0,1)3} 
  \put(-5.6,-4.2){\line(-2,-1){3.6}}
  \put(5.6,-4.2){\line(2,-1){3.6}}
 \end{picture}
 \Longrightarrow \FR 1 3
 \begin{picture}(24,11)(-11,-1) 
  \put(0,0){\circle{14}}
  \put(0,7){\line(-1,-2){5.6}} 
  \put(0,7){\line(1,-2){5.6}}
  \put(-8.6,-4.2){\line(1,0){17.2}} 
  \put(-5.6,4.2){\vector(-1,-2){.3}} 
  \put(-2.8,1.4){\vector(-1,-2){.3}}
  \put(-4.2,2.8){\circle*{.6}}
  \put(5.6,4.2){\vector(1,-2){.3}} 
  \put(2.8,1.4){\vector(1,-2){.3}}
  \put(4.2,2.8){\circle*{.6}}
  \put(0,7){\circle*1} 
  \put(-5.6,-4.2){\circle*1}
  \put(5.6,-4.2){\circle*1}
  \put(-2,-12){\mbox (a)}
 \end{picture}
 + \FR 2 3
 \begin{picture}(24,11)(-11,-1) 
  \put(0,0){\circle{14}}
  \put(0,7){\line(-1,-2){5.6}} 
  \put(0,7){\line(1,-2){5.6}}
  \put(-8.6,-4.2){\line(1,0){17.2}} 
  \put(-5.6,4.2){\vector(1,2){.3}} 
  \put(-2.8,1.4){\vector(1,2){.3}}
  \put(-4.2,2.8){\circle*{.6}}
  \put(0,-4.2){\vector(1,0){.7}} 
  \put(0,-7){\vector(1,0){.7}}
  \put(0,-5.6){\circle*{.6}}
  \put(0,7){\circle*1} 
  \put(-5.6,-4.2){\circle*1}
  \put(5.6,-4.2){\circle*1}
  \put(-2,-12){\mbox (b)}
 \end{picture} \vspace{4mm}
 $$
 \caption{Two possibilities that arise when one of the external momenta
  is set to zero.} \label{1zero}

\end{figure}
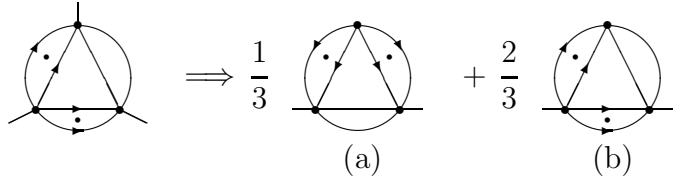

The model involves intrinsic infrared divergencies due to zero masses on
the lines without components of momenta in numerators. Besides, there
may appear artificial infrared divergencies if some of the external
momenta in fig.\ref{4loop}a are set to zero. In fact, for technical
reasons, we simply cannot do without this, to reduce the problem to
successively evaluating one-loop propagator-type graphs. The three
external lines of the diagram correspond to identical operator
structures of the background field. Therefore, nullifying one of the
three momenta, we obtain two different graphs shown in fig.\ref{1zero},
while setting all of them to zero, we get a pure tadpole. The diagram of
fig.\ref{1zero}b and the complicated tadpole graph contain artificial
infrared divergencies.

We can easily compute the two one-loop subgraphs both in co-ordinate and
in momentum space (\ref{an}) by means of eqs.(\ref{p1}), (\ref{IL}),
(\ref{il}), and (\ref{JL}):
\begin{eqnarray} &0
 \begin{picture}(10,5)(-5,-1) 
  \put(0,0){\circle 7}
  \put(-3.5,0){\circle*1}
  \put(3.5,0){\circle*1}
  \put(0,3.5){\vector(1,0){.7}}
  \put(0,-3.5){\vector(1,0){.7}}
  \put(0,0){\circle*{.6}}
 \end{picture}
 x~ \Longrightarrow~ - \FR 1 {4 \pi^2} \FR 1 {x^2} , \s2
 \begin{picture}(17,5)(-7,-1) 
  \put(0,0){\circle 7}
  \put(-3.5,0){\circle*1}
  \put(3.5,0){\circle*1}
  \put(0,3.5){\vector(1,0){.7}}
  \put(0,-3.5){\vector(1,0){.7}}
  \put(0,0){\circle*{.6}}
  \put(-3.5,0){\line(-1,0)3}
  \put(3.5,0){\line(1,0)3}
  \put(5,2){\mbox{$p$}}
 \end{picture}
 \Longrightarrow \FR 1 {4\pi} \ln (p^2/\mu^2)~ , \s1& \label{UV1} \\
&0
 \begin{picture}(10,5)(-5,-1) 
  \put(0,0){\circle 7}
  \put(-3.5,0){\circle*1}
  \put(3.5,0){\circle*1}
 \end{picture}
 x~ \Longrightarrow \FR 1 {(4\pi)^2} \ln^2 (x^2 \Nu^2)~ , \s1
 \begin{picture}(17,5)(-7,-1) 
  \put(0,0){\circle 7}
  \put(-3.5,0){\circle*1}
  \put(3.5,0){\circle*1}
  \put(-3.5,0){\line(-1,0)3}
  \put(3.5,0){\line(1,0)3}
  \put(5,2){\mbox{$p$}}
 \end{picture}
 \Longrightarrow \FR 1 {4\pi} \FR {2~\ln (p^2/\nu^2)} {p^2} . \s1&
 \label{IR1}
\end{eqnarray}
Now, the diagram of fig.\ref{1zero}a can be calculated as follows. We
first compute its two-loop propagator-type subgraph in momentum space as
the square of the one-loop expression (\ref{UV1}). Then, we transform
the subgraph to co-ordinate space
\begin{equation} 0
 \begin{picture}(17,5)(-5,-1) 
  \put(0,0){\circle 7}
  \put(7,0){\circle 7}
  \put(-3.5,0){\circle*1}
  \put(3.5,0){\circle*1}
  \put(10.5,0){\circle*1}
  \put(0,3.5){\vector(-1,0){.7}}
  \put(0,-3.5){\vector(-1,0){.7}}
  \put(0,0){\circle*{.6}}
  \put(7,3.5){\vector(1,0){.7}}
  \put(7,-3.5){\vector(1,0){.7}}
  \put(7,0){\circle*{.6}}
 \end{picture}
 x~ \Longrightarrow \int \FR {\d^2 p} {4\pi^2} \e^{-\i p x}
 \left[ \FR {\ln(p^2/\mu^2)} {4\pi}
 \right] ^{\displaystyle 2}
 = \FR 1 {\pi(4\pi)^2} \FR {2~\ln(x^2\Mu^2)} {x^2} . \label{UV2}
\end{equation}
The whole diagram is just a product of eq.(\ref{UV2}) and the infrared
one-loop subgraph (\ref{IR1}) in $x$ space. Re-expanding the infrared
logarithm squared in terms of $\ln(x^2\Mu^2)$, and switching back to
momentum space by means of eq.(\ref{I=}), we arrive at the result

\begin{eqnarray} &{\rm fig.\ref{1zero}a}~ \Rightarrow
 \FR 1 {(4\pi)^4} \Bigl[ \Bigr. ~\fr 1 2 \ln^4 (p^2/\mu^2)
 - \fr 4 3 \ln^3 (p^2/\mu^2)~ \ln (\nu^2/\mu^2)
 + \ln^2 (p^2/\mu^2)~ \ln^2 (\nu^2/\mu^2)& \nonumber\\*
&+~8~ \zeta(3)~ \ln (p^2/\mu^2)
 - \fr{16}3 \zeta(3)~ \ln (\nu^2/\mu^2)~
 \Bigl. \Bigr] .& \label{4a}
\end{eqnarray}
The same result is obtained if we leave the one-loop subgraph as it is,
but reduce eq.(\ref{UV2}) to $\ln(x^2\Nu^2)$, take the Fourier transform
of the product via eq.(\ref{IL}), and finally, re-expand it in the
ultraviolet logarithms.

In this way, using only the Fourier transformation formulae, we can
compute any diagram of recursively one-loop propagator structure. In
particular, we consecutively evaluate
\begin{eqnarray} && 0
 \begin{picture}(17,5)(-5,-1) 
  \put(0,0){\circle 7}
  \put(7,0){\circle 7}
  \put(-3.5,0){\circle*1}
  \put(3.5,0){\circle*1}
  \put(10.5,0){\circle*1}
  \put(0,3.5){\vector(1,0){.7}}
  \put(0,-3.5){\vector(1,0){.7}}
  \put(0,0){\circle*{.6}}
 \end{picture}
 x~ \Longrightarrow \int \FR {\d^2 p} {4\pi^2} \e^{-\i p x}~
 \FR {\ln(p^2/\mu^2)} {4\pi}  \FR 1{4\pi} \FR {2~\ln(p^2/\nu^2)} {p^2}
 \label{IR2}\\*
&& \s1 = \FR 1 {(4\pi)^3}
 \Bigl[ ~\fr 2 3 \ln^3 (x^2 \Mu^2)
  + \ln^2 (x^2 \Mu^2)~ \ln (\nu^2/\mu^2)
  - \fr 1 3 \ln^3 (\nu^2/\mu^2) + \fr 8 3 \zeta(3)~
 \Bigr] , \nonumber
\end{eqnarray}
\pagebreak[1]
$$ {\rm fig.\ref{1zero}b}~ \Rightarrow
 \FR 1 {(4\pi)^4} \Bigl[ \Bigr. ~\fr 1 6 \ln^4 (p^2/\mu^2)
 - \fr 1 3 \ln^3 (p^2/\mu^2)~ \ln (\nu^2/\mu^2)
 + \fr 1 3 \ln (p^2/\mu^2)~ \ln^3 (\nu^2/\mu^2) $$
\begin{equation}
 - \fr 4 3 \zeta(3)~ \ln (\nu^2/\mu^2)~ \Bigl. \Bigr] . \label{4b}
\end{equation}

The answer for the tadpole diagram at zero momenta in fig.\ref{4loop}a
proves to be ambiguous. If we first compute its infrared-regular
two-loop subgraph (\ref{UV2}), and then evaluate the tadpole by the
$\delta$-function trick (\ref{trick}) or by eq.(\ref{tx}), we get
\begin{equation}
 \begin{picture}(21,4)(-10,-1) 
  \put(0,0){\circle{14}}
  \put(0,7){\line(-1,-2){5.6}} 
  \put(0,7){\line(1,-2){5.6}}
  \put(-5.6,-4.2){\line(1,0){11.2}}
  \put(-5.6,4.2){\vector(-1,-2){.3}} 
  \put(-2.8,1.4){\vector(-1,-2){.3}}
  \put(-4.2,2.8){\circle*{.6}}
  \put(5.6,4.2){\vector(1,-2){.3}} 
  \put(2.8,1.4){\vector(1,-2){.3}}
  \put(4.2,2.8){\circle*{.6}}
  \put(0,7){\circle*1} 
  \put(-5.6,-4.2){\circle*1}
  \put(5.6,-4.2){\circle*1}
 \end{picture} \vspace{2mm}
 \Longrightarrow \FR 1 {(4\pi)^4}
 \Bigl[ ~\fr 1 6 \ln^4 (\nu^2/\mu^2)
  + \fr 8 3 \zeta(3)~ \ln (\nu^2/\mu^2)~
 \Bigr] , \label{ta}
\end{equation}
like in eq.(\ref{X1}). The use of the direct momentum-space formula
(\ref{tp}) leads to the same answer. On the other hand, starting from
the infrared-divergent two-loop subgraph (\ref{IR2}), we repeat
eq.(\ref{X2}):
\begin{equation}
 \begin{picture}(21,4)(-10,-1) 
  \put(0,0){\circle{14}}
  \put(0,7){\line(-1,-2){5.6}} 
  \put(0,7){\line(1,-2){5.6}}
  \put(-5.6,-4.2){\line(1,0){11.2}}
  \put(-5.6,4.2){\vector(1,2){.3}} 
  \put(-2.8,1.4){\vector(1,2){.3}}
  \put(-4.2,2.8){\circle*{.6}}
  \put(0,-4.2){\vector(1,0){.7}} 
  \put(0,-7){\vector(1,0){.7}}
  \put(0,-5.6){\circle*{.6}}
  \put(0,7){\circle*1} 
  \put(-5.6,-4.2){\circle*1}
  \put(5.6,-4.2){\circle*1}
 \end{picture} \vspace{2mm}
 \Longrightarrow \FR 1 {(4\pi)^4}
 \Bigl[ ~\fr 1 6 \ln^4 (\nu^2/\mu^2)
  - \fr 4 3 \zeta(3)~ \ln (\nu^2/\mu^2)~
 \Bigr] . \label{tb}
\end{equation}
Both answers (\ref{ta}) and (\ref{tb}) can be obtained from
eqs.(\ref{4a}) and (\ref{4b}), respectively, by replacing $p^2$ with
$\nu^2$.

To estimate the meaning of the results, it is worth comparing them with
the unconditionally consistent scheme of dimensional renormalization.
The infrared $\tilde{R}$ and ultraviolet $R$ operation ought to be done
explicitly then. Without entering into details, let us present as an
example (fig.\ref{R'}) the structure of renormalizations for the diagram
of fig.\ref{1zero}b.

\begin{figure}[htbp]
\begin{eqnarray*} \tilde{R} R'
 \begin{picture}(20,8.5)(-10,-2) 
  \put(0,0){\circle{14}}
  \put(0,7){\line(-1,-2){5.6}} 
  \put(0,7){\line(1,-2){5.6}}
  \put(-8.6,-4.2){\line(1,0){17.2}} 
  \put(-5.6,4.2){\vector(1,2){.3}} 
  \put(-2.8,1.4){\vector(1,2){.3}}
  \put(-4.2,2.8){\circle*{.6}}
  \put(0,-4.2){\vector(1,0){.7}} 
  \put(0,-7){\vector(1,0){.7}}
  \put(0,-5.6){\circle*{.6}}
  \put(0,7){\circle*1} 
  \put(-5.6,-4.2){\circle*1}
  \put(5.6,-4.2){\circle*1}
 \end{picture}
 &=& \tilde{R}
 \left\{
  \begin{picture}(20,8.5)(-10,-2) 
   \put(0,0){\circle{14}}
   \put(0,7){\line(-1,-2){5.6}} 
   \put(0,7){\line(1,-2){5.6}}
   \put(-8.6,-4.2){\line(1,0){17.2}} 
   \put(-5.6,4.2){\vector(1,2){.3}} 
   \put(-2.8,1.4){\vector(1,2){.3}}
   \put(-4.2,2.8){\circle*{.6}}
   \put(0,-4.2){\vector(1,0){.7}} 
   \put(0,-7){\vector(1,0){.7}}
   \put(0,-5.6){\circle*{.6}}
   \put(0,7){\circle*1} 
   \put(-5.6,-4.2){\circle*1}
   \put(5.6,-4.2){\circle*1}
  \end{picture}
  -
  \left[ K R'
   \begin{picture}(14,5)(-7,-1) 
    \put(0,0){\circle 7}
    \put(-3.5,0){\circle*1}
    \put(3.5,0){\circle*1}
    \put(0,3.5){\vector(1,0){.7}}
    \put(0,-3.5){\vector(1,0){.7}}
    \put(0,0){\circle*{.6}}
    \put(-3.5,0){\line(-1,0)3}
    \put(3.5,0){\line(1,0)3}
   \end{picture}
  \right]
  \left[
   \begin{picture}(13,8)(-6.5,-2) 
    \put(0,0){\oval(11.2,11.2)}
    \put(0,0){\oval(4.2,11.2)}
    \put(-5.6,0){\vector(0,1){.7}} 
    \put(-2.1,0){\vector(0,1){.7}}
    \put(-3.8,0){\circle*{.6}}
    \put(0,-5.6){\circle*1} 
    \put(0,5.6){\circle*1}
    \put(0,-5.6){\line(-1,-1){3.6}} 
    \put(0,-5.6){\line(1,-1){3.6}}
   \end{picture}
   +
   \begin{picture}(17,7)(-8.5,-1) 
    \put(0,0){\oval(11.2,11.2)}
    \put(0,0){\oval(11.2,4.2)}
    \put(0,-5.6){\vector(1,0){.7}} 
    \put(0,-2.1){\vector(1,0){.7}}
    \put(0,-3.8){\circle*{.6}}
    \put(-5.6,0){\circle*1} 
    \put(5.6,0){\circle*1}
    \put(-5.6,0){\line(-1,0)2} 
    \put(5.6,0){\line(1,0)2}
   \end{picture}
  \right]
 \right.\\*
&& \s{.7}
 \left. -
  \left[ K R'
   \begin{picture}(20,5)(-6.5,-1) 
    \put(0,0){\circle 7}
    \put(7,0){\circle 7}
    \put(0,3.5){\vector(-1,0){.7}} 
    \put(0,-3.5){\vector(-1,0){.7}}
    \put(0,0){\circle*{.6}}
    \put(7,3.5){\vector(1,0){.7}} 
    \put(7,-3.5){\vector(1,0){.7}}
    \put(7,0){\circle*{.6}}
    \put(-3.5,0){\circle*1} 
    \put(3.5,0){\circle*1}
    \put(10.5,0){\circle*1}
    \put(-3.5,0){\line(-1,0)2} 
    \put(10.5,0){\line(1,0)2}
   \end{picture}
  \right]
  \begin{picture}(10,7)(-5,-1) 
   \put(0,2.8){\circle{5.6}}
   \put(0,-2.8){\circle{5.6}}
   \put(-4,0){\line(1,0)8}
   \put(0,0){\circle*1}
  \end{picture}
  - 2
  \left[ K R' \tilde{R}
   \begin{picture}(17,7)(-8.5,-1) 
    \put(0,0){\circle{11.2}}
    \put(-5.6,5.6){\oval(11.2,11.2)[rb]}
    \put(0,0){\oval(11.2,4.2)[b]}
    \put(-4,4){\vector(1,1){.5}} 
    \put(-1.6,1.6){\vector(1,1){.5}}
    \put(-2.8,2.8){\circle*{.6}}
    \put(0,-2.1){\vector(1,0){.7}} 
    \put(0,-5.6){\vector(1,0){.7}}
    \put(0,-3.8){\circle*{.6}}
    \put(0,5.6){\circle*1} 
    \put(-5.6,0){\circle*1}
    \put(5.6,0){\circle*1}
    \put(-5.6,0){\line(-1,0)2} 
    \put(5.6,0){\line(1,0)2}
   \end{picture}
  \right]
  \begin{picture}(10,8.5)(-5,-2.5) 
   \put(0,0){\circle 7}
   \put(0,-3.5){\line(-1,-1){3.6}}
   \put(0,-3.5){\line(1,-1){3.6}}
   \put(0,-3.5){\circle*1}
  \end{picture}
 \right\}
\end{eqnarray*}
$$ =
 \begin{picture}(20,8.5)(-10,-2) 
  \put(0,0){\circle{14}}
  \put(0,7){\line(-1,-2){5.6}} 
  \put(0,7){\line(1,-2){5.6}}
  \put(-8.6,-4.2){\line(1,0){17.2}} 
  \put(-5.6,4.2){\vector(1,2){.3}} 
  \put(-2.8,1.4){\vector(1,2){.3}}
  \put(-4.2,2.8){\circle*{.6}}
  \put(0,-4.2){\vector(1,0){.7}} 
  \put(0,-7){\vector(1,0){.7}}
  \put(0,-5.6){\circle*{.6}}
  \put(0,7){\circle*1} 
  \put(-5.6,-4.2){\circle*1}
  \put(5.6,-4.2){\circle*1}
 \end{picture}
 ~-~ 2
 \left[ \tilde{K} \tilde{R}'
  \begin{picture}(9,2)(-4.5,-1) 
   \put(-3,0){\line(1,0)6}
   \put(-3,0){\circle*1}
   \put(3,0){\circle*1}
  \end{picture}
 \right]
 \begin{picture}(17,7)(-8.5,-1) 
  \put(0,0){\circle{11.2}}
  \put(-5.6,5.6){\oval(11.2,11.2)[rb]}
  \put(0,0){\oval(11.2,4.2)[b]}
  \put(-4,4){\vector(1,1){.5}} 
  \put(-1.6,1.6){\vector(1,1){.5}}
  \put(-2.8,2.8){\circle*{.6}}
  \put(0,-2.1){\vector(1,0){.7}} 
  \put(0,-5.6){\vector(1,0){.7}}
  \put(0,-3.8){\circle*{.6}}
  \put(0,5.6){\circle*1} 
  \put(-5.6,0){\circle*1}
  \put(5.6,0){\circle*1}
  \put(-5.6,0){\line(-1,0)2} 
  \put(5.6,0){\line(1,0)2}
 \end{picture}
 ~-~
 \left[ \tilde{K} \tilde{R}' R
  \begin{picture}(17,5)(-5,-1) 
   \put(0,0){\circle 7}
   \put(7,0){\circle 7}
   \put(-3.5,0){\circle*1}
   \put(3.5,0){\circle*1}
   \put(10.5,0){\circle*1}
   \put(0,3.5){\vector(1,0){.7}}
   \put(0,-3.5){\vector(1,0){.7}}
   \put(0,0){\circle*{.6}}
  \end{picture}
 \right]
 \left[ (1 - K R')
  \begin{picture}(14,5)(-7,-1) 
   \put(0,0){\circle 7}
   \put(-3.5,0){\circle*1}
   \put(3.5,0){\circle*1}
   \put(0,3.5){\vector(1,0){.7}}
   \put(0,-3.5){\vector(1,0){.7}}
   \put(0,0){\circle*{.6}}
   \put(-3.5,0){\line(-1,0)3}
   \put(3.5,0){\line(1,0)3}
  \end{picture}
 \right] $$ \vspace{1mm}
$$ +~
 \left[ K R'
  \begin{picture}(14,5)(-7,-1) 
   \put(0,0){\circle 7}
   \put(-3.5,0){\circle*1}
   \put(3.5,0){\circle*1}
   \put(0,3.5){\vector(1,0){.7}}
   \put(0,-3.5){\vector(1,0){.7}}
   \put(0,0){\circle*{.6}}
   \put(-3.5,0){\line(-1,0)3}
   \put(3.5,0){\line(1,0)3}
  \end{picture}
 \right]
 \left\{ ~2
  \left[ \tilde{K} \tilde{R}'
   \begin{picture}(9,2)(-4.5,-1) 
    \put(-3,0){\line(1,0)6}
    \put(-3,0){\circle*1}
    \put(3,0){\circle*1}
   \end{picture}
  \right]
  \begin{picture}(16,6)(-8,-1) 
   \put(0,0){\circle{9.8}}
   \put(-7,0){\line(1,0){14}}
   \put(-4.9,0){\circle*1}
   \put(4.9,0){\circle*1}
   \put(0,-4.9){\vector(1,0){.7}} 
   \put(0,0){\vector(1,0){.7}}
   \put(0,-2.4){\circle*{.6}}
  \end{picture}
  +
  \left[ \tilde{K} \tilde{R}'
   \begin{picture}(10,5)(-5,-1) 
    \put(0,0){\circle 7}
    \put(-3.5,0){\circle*1}
    \put(3.5,0){\circle*1}
   \end{picture}
  \right]
  \begin{picture}(14,5)(-7,-1) 
   \put(0,0){\circle 7}
   \put(-3.5,0){\circle*1}
   \put(3.5,0){\circle*1}
   \put(0,3.5){\vector(1,0){.7}}
   \put(0,-3.5){\vector(1,0){.7}}
   \put(0,0){\circle*{.6}}
   \put(-3.5,0){\line(-1,0)3}
   \put(3.5,0){\line(1,0)3}
  \end{picture} ~
 \right\} $$
$$ -~
 \left[ K R'
  \begin{picture}(20,5)(-6.5,-1) 
   \put(0,0){\circle 7}
   \put(7,0){\circle 7}
   \put(0,3.5){\vector(-1,0){.7}} 
   \put(0,-3.5){\vector(-1,0){.7}}
   \put(0,0){\circle*{.6}}
   \put(7,3.5){\vector(1,0){.7}} 
   \put(7,-3.5){\vector(1,0){.7}}
   \put(7,0){\circle*{.6}}
   \put(-3.5,0){\circle*1} 
   \put(3.5,0){\circle*1}
   \put(10.5,0){\circle*1}
   \put(-3.5,0){\line(-1,0)2} 
   \put(10.5,0){\line(1,0)2}
  \end{picture}
 \right]
 \left[ \tilde{K} \tilde{R}'
  \begin{picture}(9,2)(-4.5,-1) 
   \put(-3,0){\line(1,0)6}
   \put(-3,0){\circle*1}
   \put(3,0){\circle*1}
  \end{picture}
 \right]^2
 ~+~ 2
 \left[ K R' \tilde{R}
  \begin{picture}(17,7)(-8.5,-1) 
   \put(0,0){\circle{11.2}}
   \put(-5.6,5.6){\oval(11.2,11.2)[rb]}
   \put(0,0){\oval(11.2,4.2)[b]}
   \put(-4,4){\vector(1,1){.5}} 
   \put(-1.6,1.6){\vector(1,1){.5}}
   \put(-2.8,2.8){\circle*{.6}}
   \put(0,-2.1){\vector(1,0){.7}} 
   \put(0,-5.6){\vector(1,0){.7}}
   \put(0,-3.8){\circle*{.6}}
   \put(0,5.6){\circle*1} 
   \put(-5.6,0){\circle*1}
   \put(5.6,0){\circle*1}
   \put(-5.6,0){\line(-1,0)2} 
   \put(5.6,0){\line(1,0)2}
  \end{picture}
 \right]
 \left[ \tilde{K} \tilde{R}'
  \begin{picture}(9,2)(-4.5,-1) 
   \put(-3,0){\line(1,0)6}
   \put(-3,0){\circle*1}
   \put(3,0){\circle*1}
  \end{picture}
 \right] $$

 \caption{The structure of the incomplete ultraviolet $R'$ operation and
  the full infrared $\tilde{R}$ operation. A subtracted ultraviolet
  subgraph is shrunk to a point, while an infrared subgraph is just
  deleted. After the $\tilde{R}$, we have omitted zero tadpole graphs.}

 \label{R'}
\end{figure}

The subtraction operators $K$ and $\tilde{K}$ pick out poles in
$\varepsilon$=(2$-$$D$)/2 in momentum and co-ordinate space,
respectively. We keep different ultraviolet and infrared renormalization
scales. Thus, every momentum loop integration is, as usual, accompanied
by a $\left(\mu^2\right)^\varepsilon$ factor, while the proper dimension
of infrared counterterms proportional to $\delta_{2-2\varepsilon}(p)$
is restored by additional denominators of
$\left(\nu^2\right)^\varepsilon$. For example, in fig.\ref{R'} we should
substitute
$$ \tilde{K} \tilde{R}' R
 \begin{picture}(17,5)(-5,-1) 
  \put(0,0){\circle 7}
  \put(7,0){\circle 7}
  \put(-3.5,0){\circle*1}
  \put(3.5,0){\circle*1}
  \put(10.5,0){\circle*1}
  \put(0,3.5){\vector(1,0){.7}}
  \put(0,-3.5){\vector(1,0){.7}}
  \put(0,0){\circle*{.6}}
 \end{picture}
 = \FR 1 {3\varepsilon^3} \left(\mu^2/\nu^2\right)^{3\varepsilon} , $$
one $\left(\mu^2\right)^\varepsilon$ coming from the integration
annihilated by the $\delta$ function.

Instead of doing the infrared $\tilde{R}$ operation in the diagram
language, as in fig.\ref{R'}, it may technically be easier to subtract
singular momentum-space expressions directly by means of a simple
formula \cite{sig5}
$$ \FR 1 {\left(p^2\right)^{1+n\varepsilon}}
 \stackrel {\strut\displaystyle\tilde{R}} \to
 \FR 1 {\left(p^2\right)^{1+n\varepsilon}}
 + \pi \FR {\left(\nu^2\right)^{-(n+1)\varepsilon}} {(n+1)~\varepsilon}
  \delta_{2-2\varepsilon} (p)~, $$
leaving only pole terms on subtracting complicated subgraphs with a
prefactor.

\pagebreak[1] Here are the results of dimensional renormalization in the
so-called $G$-scheme \cite{G-}
\begin{eqnarray} (4\pi)^4 &R' \tilde{R}&
 \begin{picture}(20,8)(-10,-2) 
  \put(0,0){\circle{14}}
  \put(0,7){\line(-1,-2){5.6}} 
  \put(0,7){\line(1,-2){5.6}}
  \put(-8.6,-4.2){\line(1,0){17.2}} 
  \put(-5.6,4.2){\vector(-1,-2){.3}} 
  \put(-2.8,1.4){\vector(-1,-2){.3}}
  \put(-4.2,2.8){\circle*{.6}}
  \put(5.6,4.2){\vector(1,-2){.3}} 
  \put(2.8,1.4){\vector(1,-2){.3}}
  \put(4.2,2.8){\circle*{.6}}
  \put(0,7){\circle*1} 
  \put(-5.6,-4.2){\circle*1}
  \put(5.6,-4.2){\circle*1}
 \end{picture} \vspace{4mm}
 = - \FR 1 {6\varepsilon^4} + \FR {\zeta(3)} \varepsilon
 \nonumber\\*
&& + \fr 1 2 \ln^4 (p^2/\mu^2)
 - \fr 4 3 \ln^3 (p^2/\mu^2)~ \ln (\nu^2/\mu^2)
 + \ln^2 (p^2/\mu^2)~ \ln^2 (\nu^2/\mu^2) \nonumber\\*
&& +~ 8 \zeta(3)~ \ln (p^2/\mu^2)
 - \fr{28}3 \zeta(3)~ \ln (\nu^2/\mu^2) + \fr 3 2 \zeta(4)~, \label{4A}
 \\
(4\pi)^4 &R' \tilde{R}&
 \begin{picture}(20,9)(-10,-2) 
  \put(0,0){\circle{14}}
  \put(0,7){\line(-1,-2){5.6}} 
  \put(0,7){\line(1,-2){5.6}}
  \put(-8.6,-4.2){\line(1,0){17.2}} 
  \put(-5.6,4.2){\vector(1,2){.3}} 
  \put(-2.8,1.4){\vector(1,2){.3}}
  \put(-4.2,2.8){\circle*{.6}}
  \put(0,-4.2){\vector(1,0){.7}} 
  \put(0,-7){\vector(1,0){.7}}
  \put(0,-5.6){\circle*{.6}}
  \put(0,7){\circle*1} 
  \put(-5.6,-4.2){\circle*1}
  \put(5.6,-4.2){\circle*1}
 \end{picture} \vspace{5mm}
 = - \FR 1 {6\varepsilon^4} + \FR {\zeta(3)} \varepsilon \nonumber\\*
&& + \fr 1 6 \ln^4 (p^2/\mu^2)
 - \fr 1 3 \ln^3 (p^2/\mu^2)~ \ln (\nu^2/\mu^2)
 + \fr 1 3 \ln (p^2/\mu^2)~ \ln^3 (\nu^2/\mu^2) \nonumber\\*
&& - \fr 4 3 \zeta(3)
 \Bigl[ \ln (p^2/\mu^2) + \ln (\nu^2/\mu^2)
 \Bigr] + \fr 3 2 \zeta(4)~, \label{4B} \\
(4\pi)^4 &R' \tilde{R}&
 \begin{picture}(20,8)(-10,-1) 
  \put(0,0){\circle{14}}
  \put(0,7){\line(-1,-2){5.6}} 
  \put(0,7){\line(1,-2){5.6}}
  \put(-5.6,-4.2){\line(1,0){11.2}}
  \put(-5.6,4.2){\vector(-1,-2){.3}} 
  \put(-2.8,1.4){\vector(-1,-2){.3}}
  \put(-4.2,2.8){\circle*{.6}}
  \put(5.6,4.2){\vector(1,-2){.3}} 
  \put(2.8,1.4){\vector(1,-2){.3}}
  \put(4.2,2.8){\circle*{.6}}
  \put(0,7){\circle*1} 
  \put(-5.6,-4.2){\circle*1}
  \put(5.6,-4.2){\circle*1}
 \end{picture} \vspace{4mm}
 = - \FR 1 {6\varepsilon^4} + \FR {\zeta(3)} \varepsilon
 + \fr 1 6 \ln^4 (\nu^2/\mu^2) - 4 \zeta(3)~ \ln (\nu^2/\mu^2)~. \s1
 \label{4T}
\end{eqnarray}

Comparing eqs.(\ref{4a})--(\ref{tb}) with eqs.(\ref{4A})--(\ref{4T}),
we come to the following conclusions. The results of the extended
differential renormalization disagree with the dimensionally
renormalized diagrams obtained by subtracting pole contributions,
$R\tilde{R}$=(1$-$$K)R'\tilde{R}$. Moreover, this disagreement does not
fit into the framework of the scheme arbitrariness by just redefining
$\mu$ and $\nu$: eq.(\ref{4B}) involves a $\ln(p^2/\mu^2)$ term absent
in eq.(\ref{4b}). Forming the weighed sum, according to fig.\ref{1zero},
does not help in improving the agreement. The tadpole graphs are not
well-defined at all, none of the results coinciding with eq.(\ref{4T}),
neither eq.(\ref{ta}) nor (\ref{tb}).

One cannot completely avoid tadpole-graph ambiguities even at
$\mu$=$\nu$ (this would be sufficient for asymptotic expansions).
Already for the five-loop tadpole diagram shown in fig.\ref{tadpole5},
we can get three different answers at $\mu$=$\nu$, starting from its
one-, two-, or three-loop infrared-divergent subgraphs. This is not just
an artificial example: the graph appears in the five-loop calculations
of ref.\cite{sig5}.

\begin{figure}[htbp] $$ \left( 4 \pi \right)^5
 \begin{picture}(18,7.5)(-9,-1) 
  \put(0,0){\circle{14}}
  \put(-7,7){\oval(14,14)[rb]}
  \put(7,7){\oval(14,14)[lb]}
  \put(-7,-7){\oval(14,14)[rt]}
  \put(7,-7){\oval(14,14)[lt]}
  \put(0,7){\circle*1}
  \put(0,-7){\circle*1}
  \put(-7,0){\circle*1}
  \put(7,0){\circle*1}
  \put(4.95,4.95){\vector(-1,1){.5}}
  \put(2.05,2.05){\vector(-1,1){.5}}
  \put(3.5,3.5){\circle*{.6}}
  \put(-4.95,-4.95){\vector(1,-1){.5}}
  \put(-2.05,-2.05){\vector(1,-1){.5}}
  \put(-3.5,-3.5){\circle*{.6}}
  \put(4.95,-4.95){\vector(1,1){.5}}
  \put(2.05,-2.05){\vector(1,1){.5}}
  \put(3.5,-3.5){\circle*{.6}}
 \end{picture}
 \Longrightarrow \FR{48}5 \zeta(5),~ \FR{32}5 \zeta(5),~
 \FR{12}5 \zeta(5). $$
 \caption{The five-loop tadpole graph, the value of which in the
  extended differential renormalization is ambiguous even at
  $\mu$=$\nu$.}
 \label{tadpole5}
\end{figure}

Trying to extract from eqs.(\ref{4a})--(\ref{tb}) the contributions to
the renormalization-group $\beta$ function by differentiating the
expressions with respect to $\ln\mu^2$, we find that the result depends
on $\ln(\nu^2/\mu^2)$ and $\ln(p^2/\mu^2)$. In the present case, the
only nontrivial diagram of fig.\ref{4loop}a is proportional to a new
invariant structure in the Riemann tensors and the background field,
which was not present in lower orders of perturbation theory. Thus, no
compensation of higher-order logarithms from lower-order graphs can
occur, as it happens in ordinary renormalizable field theories. The
ignored contributions from the primitive tadpole-reducible graphs of
fig.\ref{4loop}b cannot save the situation with higher-order logarithms
of the momentum since they contain nothing but $\ln^4(\nu^2/\mu^2)$.
This problem is present in dimensional renormalization as well, if we
attempt to find the $\beta$ function by using the finite logarithms
instead of applying the traditional method \cite{MS} based on the
first-order pole in $\varepsilon$. Even at $p^2$=$\nu^2$=$\mu^2$,
fig.\ref{1zero} with the finite parts of eqs.(\ref{4A}) and (\ref{4B})
does not reproduce the correct $\beta$ function \cite{sig2} extracted
from the pole term, although the dimensionally renormalized tadpole
graph gives the consistent answer in both ways. The cause of these
difficulties is non-renormalizability of the model, as a consequence of
which the renormalization-group-like equations can only be written in
terms of the pole contributions rather than directly in finite
logarithms \cite{GenRG,sig1}.

\subsection{A renormalizable sigma model} \label{ren} \indent

It would be natural then to consider a renormalizable model, where the
re\-nor\-mal\-iza\-tion-group functions can as well be computed from
finite logarithms. As such a model, we can choose the $n$-component
$\vec{\bf n}$ field in two dimensions,
$$ {\cal L} = ( 2 h )^{-1}
 \left( \partial_\mu \vec{\bf n}
 \right)^2, \s2 \vec{\bf n}^2=1~, $$
which can be rewritten as a special case of the simplest bosonic
$\sigma$ model \cite{sig1} for $n$$-$1 independent components
$$ {\cal L} = \fr 1 2 (\partial_\mu \phi^j)~ g_{jk}~
 \partial_\mu \phi^k, \s2
 g_{jk} = \delta_{jk} + \FR h {1-h\phi^2} \phi_j \phi_k~ . $$
For this particular form of the metric with the co-ordinates $\phi_j$,
we have
$$ g^{jk} = \delta^{jk} - h~ \phi^j \phi^k, \s2
 {\Gamma_{jk}}^l = h~ g_{jk}~ \phi^l~,$$
$$ R_{jklm} = h
 \left( g_{jl}~ g_{km} - g_{jm}~ g_{kl}
 \right) , \s2 R_{jk} = (n-2)~ h~ g_{jk}~ . $$
The invariant charge $\overline{h}$=$Z^{-1}h$ is determined by the
charge renormalization constant $Z$. The contributing Feynman diagrams
in the background-field formalism \cite{sig1} up to two loops are
presented in fig.\ref{Z2}.

\begin{figure}[htbp]
 $$ - R_{jk}~ (\partial_\mu \phi^j)~ \partial_\mu \phi^k~
  \begin{picture}(10,6)(-5,-2.5) 
   \put(0,0){\circle 7}
   \put(0,-3.5){\line(-1,-1){3.6}}
   \put(0,-3.5){\line(1,-1){3.6}}
   \put(0,-3.5){\circle*1}
  \end{picture} \eqnum{\rm a} $$
 $$ + \fr 1{12}
  \left( 2~ {R^a}_j~ R_{ak} + 3~ {R^{abc}}_j~ R_{abck}
  \right) (\partial_\mu \phi^j)~ \partial_\mu \phi^k~
  \begin{picture}(10,7)(-5,-1) 
   \put(0,2.8){\circle{5.6}}
   \put(0,-2.8){\circle{5.6}}
   \put(-4,0){\line(1,0)8}
   \put(0,0){\circle*1}
  \end{picture} \eqnum{\rm b} $$
 $$ - \fr 1 6 {R_j}^{ab}{}_k~ R_{ab}~
  (\partial_\mu \phi^j)~ \partial_\mu \phi^k~
  \begin{picture}(10,8)(-5,-1) 
   \put(0,0){\circle{5.6}}
   \put(0,5.6){\circle{5.6}}
   \put(-2.8,0){\vector(0,-1){.7}} 
   \put(2.8,0){\vector(0,-1){.7}}
   \put(0,0){\circle*{.6}}
   \put(0,2.8){\circle*1} 
   \put(0,-2.8){\circle*1}
   \put(0,-2.8){\line(-1,-1){3.6}} 
   \put(0,-2.8){\line(1,-1){3.6}}
  \end{picture} \eqnum{\rm c} $$
 $$ + \fr 4 9 {R_j}^{(ab)c}~ R_{kabc}~
  (\partial_\mu \phi^j)~ \partial_\nu \phi^k~
  \begin{picture}(16,9)(-8,-1) 
   \put(0,0){\circle{9.8}}
   \put(-7,0){\line(1,0){14}}
   \put(-4.9,0){\circle*1}
   \put(4.9,0){\circle*1}
   \put(-2.5,0){\vector(1,0){.7}} 
   \put(2.5,0){\vector(-1,0){.7}}
  \end{picture} \eqnum{\rm d} $$
 $$ + \fr 8 9 {R_j}^{(ab)c}~ R_{k(cb)a}~
  (\partial_\mu \phi^j)~ \partial_\nu \phi^k~
  \begin{picture}(16,8)(-8,-1) 
   \put(0,0){\circle{9.8}}
   \put(-7,0){\line(1,0){14}}
   \put(-4.9,0){\circle*1}
   \put(4.9,0){\circle*1}
   \put(-3.5,3.5){\vector(1,1){.5}} 
   \put(3.5,-3.5){\vector(-1,-1){.5}}
  \end{picture} \eqnum{\rm e} $$

 \caption{The one- and two-loop corrections to the effective action of
  the two-dimensional bosonic $\sigma$ model without torsion. The
  contributions to $Z$ are obtained by normalizing to the tree term
  $\fr{1}2 (\partial_\mu \phi^j)~ g_{jk}~ \partial_\mu \phi^k$.}

 \label{Z2}
\end{figure}

\noindent The results of dimensional renormalization for the
$\vec{\bf n}$ field are
\begin{eqnarray}
R' \tilde{R}~ {\rm(fig.\ref{Z2}a)} &=& - (n-2)~
 g_{jk}~ (\partial_\mu \phi^j) (\partial_\mu \phi^k) \FR h {4\pi}
 \left[ \FR 1 \varepsilon + \ln \FR {\mu^2} {\nu^2}
 \right] , \label{2A} \\
R' \tilde{R}~ {\rm(fig.\ref{Z2}b)} &=& \fr 1 6 (n+1) (n-2)~
 g_{jk}~ (\partial_\mu \phi^j) (\partial_\mu \phi^k)
 \FR {h^2} {(4\pi)^2}
 \left[ - \FR 1 {\varepsilon^2} + \ln^2 \FR {\mu^2} {\nu^2}
 \right] ,~~~~~ \label{2B} \\
R' \tilde{R}~ {\rm(fig.\ref{Z2}c)} &=& \fr 1 6 (n-2)^2
 g_{jk}~ (\partial_\mu \phi^j) (\partial_\mu \phi^k)
 \FR {h^2} {(4\pi)^2}
 \left[ \FR 1 {\varepsilon^2} - \ln^2 \FR {\mu^2} {\nu^2}
 \right] , \label{2C}
\end{eqnarray}

\pagebreak[1]
\begin{equation} R' \tilde{R}~ {\rm(fig.\ref{Z2}d)} ~=~
 \fr 2 3 (n-2)~ g_{jk}~ (\partial_\mu \phi^j) (\partial_\nu \phi^k)
 \FR {h^2} {(4\pi)^2}
 \left[ \FR {g_{\mu\nu}} 2
  \left( \FR 1 {\varepsilon^2} - \FR 1 \varepsilon
  \right.
 \right. \s2 \label{2D}
\end{equation}
$$
 \left.
  \left. +~ \ln^2 \FR {p^2} {\nu^2} - 2 \ln \FR {p^2} {\nu^2}
   - \ln^2 \FR {\mu^2} {\nu^2} - 2 \ln \FR {\mu^2} {\nu^2} + 1
  \right)
  - \FR {p_\mu p_\nu} {p^2}
  \left( \ln^2 \FR {p^2} {\nu^2} - 2 \ln \FR {p^2} {\nu^2} + 2
  \right)
 \right] , $$

\pagebreak[1]
\begin{equation} R' \tilde{R}~ {\rm(fig.\ref{Z2}e)} ~=~
 - \fr 2 3 (n-2)~ g_{jk}~ (\partial_\mu \phi^j) (\partial_\nu \phi^k)
 \FR {h^2} {(4\pi)^2}
 \left[ - \FR {g_{\mu\nu}} 4
  \left( \FR 1 {\varepsilon^2} - \FR 1 \varepsilon
  \right.
 \right. \s2 \label{2E}
\end{equation}
$$
 \left.
  \left. +~ \ln^2 \FR {p^2} {\nu^2} - 2 \ln \FR {p^2} {\nu^2}
   - \ln^2 \FR {\mu^2} {\nu^2} - 2 \ln \FR {\mu^2} {\nu^2} + 1
  \right)
  - \FR {p_\mu p_\nu} {p^2}
  \left( \ln \FR {p^2} {\nu^2} - 1
  \right)
 \right] . $$
These formulae lead to the renormalized
$$ Z_{\rm dim} = 1 - (n-2)
 \left( 2 \ln \FR {\mu^2} {\nu^2}
 \right) \FR h {4\pi} - (n-2)
 \left( \fr 1 3 \ln^2 \FR {p^2} {\nu^2} + 2 \ln \FR {\mu^2} {p^2} + 3
 \right) \FR {h^2} {(4\pi)^2} . $$
Independence of the invariant charge on $\mu^2$ leads then to the
following expression for the $\beta$ function
\begin{eqnarray} \beta_{\rm dim} (h) = h
 \left( \FR {\partial Z} {\partial \ln \mu^2}
 \right)
 \left[
  \left( 1 - h \FR \partial {\partial h}
  \right) Z
 \right] ^{\displaystyle -1} = - 2 (n-2) \FR {h^2} {4\pi}
 \left( 1 + \FR h {4\pi}
 \right) . \label{Beta2}
\end{eqnarray}
The result is the same as obtained from the $1/\varepsilon$-pole terms
in eqs.(\ref{2A})--(\ref{2E}). It also agrees with the known $\beta$
function for the $\vec{\bf n}$-field $O(n)$ sigma model \cite{n-field}
where no infrared $\tilde{R}$ operation was ever used.


To be capable of evaluating the Feynman integrals of fig.\ref{Z2}(d,e)
in the extended differential renormalization, we need to extend the set
of formulae to a more general kind of expressions. The new differential
relation is
\begin{equation} \FR {x_\mu} {\left( x^2 \right)^2} \ln^n (x^2 \Mu^2)
 = - \fr 1 2 n! \FR \partial {\partial x_\mu}
 \left[ \FR 1 {x^2} \sum_{m=0}^n \FR {\ln^m (x^2 \Mu^2)} {m!}
  + \pi \delta_2 (x)
 \right] . \label{x3}
\end{equation}
The scale parameter $\Mu$ is the same as in eq.(\ref{x2}), while the
coefficient of the $\delta$ function is uniquely fixed by consistency
with the previously established formulae. We multiply eq.(\ref{x3}) by
$x_\mu$ and take the Fourier transform. After integrating the derivative
by parts on the right-hand side, we get the integrals of the known
types, eqs.(\ref{I}) and (\ref{x1}), while the left-hand side becomes
just $I_n$, eq.(\ref{I=}). Thus, we can easily find the necessary
coefficient.

The Feynman integrals in the extended differential renormalization, to
be compared with the expressions in square brackets in
eqs.(\ref{2A})--(\ref{2E}), can now be evaluated to
\begin{eqnarray} {\rm fig.\ref{Z2}a} &\Rightarrow& \ln (\mu^2/\nu^2)~ ,
 \label{2a} \\
{\rm fig.\ref{Z2}b} &\Rightarrow& \ln^2 (\mu^2/\nu^2)~ , \label{2b} \\
{\rm fig.\ref{Z2}c} &\Rightarrow& - \ln^2 (\mu^2/\nu^2)~ , \label{2c} \\
{\rm fig.\ref{Z2}d} &\Rightarrow& \fr 1 2 g_{\mu\nu}
 \left[ \ln^2 (p^2/\nu^2) - 2 \ln (p^2/\nu^2) - \ln^2 (\mu^2/\nu^2) + 2
 \right] \nonumber\\*
&& -~ (p_\mu p_\nu /p^2)
 \left[ \ln^2 (p^2/\nu^2) - 2 \ln (p^2/\nu^2) + 2
 \right] , \label{2d} \\
{\rm fig.\ref{Z2}e} &\Rightarrow& - \fr 1 4 g_{\mu\nu}
 \left[ \ln^2 (p^2/\nu^2) - 2 \ln (p^2/\nu^2) - \ln^2 (\mu^2/\nu^2) + 2
 \right] \nonumber\\*
&& -~ (p_\mu p_\nu /p^2) \left[ \ln (p^2/\nu^2) - 1 \right] . \label{2e}
\end{eqnarray}
We see that eqs.(\ref{2d}) and (\ref{2e}) differ from the finite parts
of eqs.(\ref{2D}) and (\ref{2E}) in dimensional renormalization by the
lack of the lower-degree logarithm of $\mu^2$. As a result, we get quite
a different charge renormalization
$$ Z_{\rm diff} = 1 - (n-2)
 \left( 2 \ln \FR {\mu^2} {\nu^2}
 \right) \FR h {4\pi} + (n-2)
 \left( - \fr 1 3 \ln^2 \FR {p^2} {\nu^2} + 2 \ln \FR {p^2} {\nu^2} - 2
 \right) \FR {h^2} {(4\pi)^2} , $$
\begin{equation} \beta_{\rm diff} (h) = - 2 (n-2)~ h^2 /(4\pi)
 + 0 \cdot h^3 . \label{beta2}
\end{equation}
Equations (\ref{beta2}) and (\ref{Beta2}) disagree beyond the range
allowed by the scheme arbitrariness in perturbation theory because in a
one-charge model the two-loop $\beta$ function must be
scheme-independent \cite{scheme2}.

Thus, the infrared extension of differential renormalization completely
fails in two-dimensional models with intrinsic infrared divergencies.
The results of multiloop renormalization-group calculations may prove to
be incorrect or ill-defined. In higher orders of perturbation theory,
ambiguities cannot be avoided even if we use the same scale $\mu$=$\nu$
for renormalizing infrared and ultraviolet divergencies.

\section{D = 4: Artificial infrared divergences} \indent

There are no off-shell infrared divergencies in $D$=4. However, they may
appear as a result of the infrared rearrangement in Feynman diagrams,
aimed to the simplification of multiloop calculations \cite{IR-re},
without which some complicated diagrams may technically occur
unattainable for analytic computation. Indeed, the usual trick is to put
some (or all) external momenta to be zero (that is  to integrate over
the corresponding co-ordinates in $x$ space), and then to calculate
divergencies, for example, the poles in dimensional regularization.
According to the theorem proven in ref.\cite{Speer}, after subtraction
of subdivergencies, the singular part of a diagram is a polynomial in
external momenta and internal masses, which is reduced to a constant
independent of them in the case of a logarithmic divergency. This allows
one to simplify calculations, taking care of the infrared divergencies
that may appear as a result of nullifying some momenta. These {\em
artificial} infrared divergencies can be subtracted by means of the
infrared $\tilde{R}$ operation \cite{R*}.

Our aim here is to check whether the formalism of differential
renormalization can be extended to perform the infrared $\tilde{R}$
operation in the same way as it performs the usual ultraviolet $R$
operation in a renormalizable model.

To be concrete, consider the massless scalar field theory
$\phi^4_{_{(D=4)}}$. The basic equation of differential renormalization
\cite{Dif} is the four-dimensional analog of eq.(\ref{x2}),
\begin{equation} 1/\left(x^2\right)^2 = - \fr 1 4 \Box_x
 \left[ \ln(x^2\Mu^2)/x^2 \right] , \label{31}
\end{equation}
and its counterpart in momentum space is
\begin{equation} 1/\left(p^2\right)^2 = - \fr 1 4 \Box_p
 \left[ \ln (p^2/\nu^2)/p^2 \right] . \label{32}
\end{equation}
Here, $\Mu$ and $\nu$ are the ultraviolet and infrared scales,
respectively.

Following the rules of the differential renormalization method
\cite{Dif}, we perform the calculations, replacing all the singular
expressions according to eqs.(\ref{31}) and (\ref{32}), and then
integrating the derivatives by parts and ignoring surface terms.
However, we should point out an important difference from our line of
reasoning with the set of Fourier-transformation formulae in two
dimensions. If we allow us to differentiate Feynman integrals,
generating {\em singular} expressions, then the circle of checking
consistency of the integrals does {\em not} close by a nonzero additive
constant. Thus, we are forced to follow a more cautious way, avoiding
any explicit differentiation, the result of which may be singular.
Neither shall we attribute any meaning to the logarithm at its singular
zero point.

Obeying these rules and using an intermediate analytic regularization
when needed, we get the following set of formulae:

\begin{eqnarray} I_n &=& \int \e^{\i p x} \d^4 x
 \FR {\ln^n (x^2\Mu^2)} {x^2} = 4\pi^2 \FR {(-)^n} {p^2}
 \FR {\d^n} {d\alpha^n}
 \left.
  \left[ \left( p^2/\mu^2 \right)^\alpha F(\alpha)
  \right]
 \right|_{\alpha = 0}~ ,\label{33} \\
i_n &=& \int \e^{-\i p x} \d^4 p \FR {\ln^n (p^2/\nu^2)} {p^2}
 = 4\pi^2 \FR {(-)^n} {x^2} \FR {\d^n} {d\alpha^n}
 \left.
  \left[
   \left(x^2\Nu^2
   \right)^\alpha F(\alpha)
  \right]
 \right|_{\alpha = 0}~ ,\label{34} \\
J_n &=& \int \e^{\i p x} \d^4 x
 \FR {\ln^n (x^2\Mu^2)} {\left(x^2\right)^2}
 = \pi^2 \FR {(-)^{n+1}~(n-1)!} {n+1} \times \label{35} \\*
&\times& \sum_{k=1}^{n} \FR 1 {(k-1)!}
 \left.
  \left\{ \FR {\d^{k+1}} {d\alpha^{k+1}}
   \left[
    \left(\FR {p^2} {\mu^2}
    \right)^\alpha F(\alpha)
   \right] - 2 k \FR {\d^k} {d\alpha^k}
   \left[
    \left( \FR {p^2} {\mu^2}
    \right)^\alpha \FR {F(\alpha)} {1+\alpha}
   \right]
  \right\}
 \right|_{\alpha = 0}~ , \nonumber \\[1mm]
j_n &=& \int \e^{-\i p x} \d^4 p
 \FR {\ln^n (p^2/\nu^2)} {\left(p^2\right)^2}
 = \pi^2 \FR {(-)^{n+1}~(n-1)!} {n+1} \times \label{36} \\*
&\times& \sum_{k=1}^{n} \FR {1} {(k-1)!}
 \left.
  \left\{ \FR {\d^{k+1}} {d\alpha^{k+1}}
   \left[ \left( x^2\Nu^2 \right)^\alpha F(\alpha)
   \right] - 2 k \FR {\d^k} {d\alpha^k}
   \left[ \left( x^2\Nu^2 \right)^\alpha \FR {F(\alpha)} {1+\alpha}
   \right]
  \right\}
 \right|_{\alpha = 0}~ , \nonumber
\end{eqnarray}
where $F(\alpha)$ is the same as in eq.(\ref{Falpha}) above. In
particular, to the lowest orders, one gets
\begin{eqnarray} I_0 &=& 4\pi^2 /p^2 , \label{37} \\
I_1 &=& - (4\pi^2/p^2)~ \ln (p^2/\mu^2)~ , \label{38} \\
I_2 &=& (4\pi^2/p^2)~ \ln^2 (p^2/\mu^2)~ , \label{39} \\
J_0 &=& - \pi^2 \ln (p^2/\mu^2)~ , \label{310} \\
J_1 &=& \pi^2
 \left[ \fr 1 2 \ln^2 (p^2/\mu^2) - \ln (p^2/\mu^2) + 1
 \right] , \label{311} \\
J_2 &=& \pi^2
 \left[ - \fr {1} {3} \ln^3 (p^2/\mu^2) + \ln^2 (p^2/\mu^2)
  - 2\ln (p^2/\mu^2) - \fr {4} {3} \zeta(3) + 2
 \right] , \label{312} \end{eqnarray}
and the same for the space-time integrals $i_n$ and $j_n$ with the
interchange $x^2$$\leftrightarrow$$p^2$,
$\Mu^2$$\leftrightarrow$$1/\nu^2$, $\mu^2$$\leftrightarrow$$1/\Nu^2$.
The relation between $\mu$ and $\Mu$ remains the same as in two
dimensions; capitals refer to $x$ space. Using
eqs.(\ref{31})--(\ref{36}), (\ref{37})--(\ref{312}), we can perform all
the calculations.

In principle, the set of formulae can be supplemented with
renormalization-scale-independent singular equations
$$ \Box_x~ (1/x^2) = - 16~ \pi^2~ \delta_4 (x)~ , \s2
 \Box_p~ (1/p^2) = - 16~ \pi^2~ \delta_4 (p)~ . $$
However, integrating a derivative by parts, we can avoid using them
explicitly, for the sake of safety always keeping ourselves at least one
step off the singular threshold.

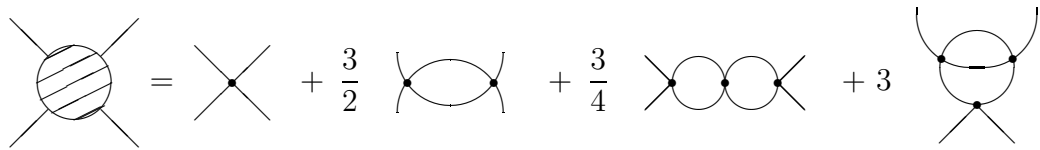
\begin{figure}[htbp] \centering
\begin{picture}(138,21)
 \put(10,10){\circle{10}}
 \put(13.54,13.54){\line(1,1)5} 
 \put(13.54,6.46){\line(1,-1)5}
 \put(6.46,13.54){\line(-1,1)5}
 \put(6.46,6.46){\line(-1,-1)5}
 \put(5.5,7.75){\line(2,1)9} 
 \put(5,10){\line(2,1)8}
 \put(15,10){\line(-2,-1)8}
 \put(10,5){\line(2,1)4}
 \put(10,15){\line(-2,-1)4}
 \put(20,9){\mbox = } 
 \put(26,5){\line(1,1){10}} 
 \put(26,15){\line(1,-1){10}}
 \put(31,10){\circle*1}
 \put(40,9){\mbox + $\FR 3 2 $} 
 \put(60,14){\oval(14.1,14.1)[b]} 
 \put(60,6){\oval(14.1,14.1)[t]}
 \put(54.25,10){\circle*1}
 \put(65.75,10){\circle*1}
 \put(73,9){\mbox + $\FR 3 4 $} 
 \put(93,10){\circle 7} 
 \put(100,10){\circle 7}
 \put(96.5,10){\circle*1}
 \put(89.5,10){\line(-1,1){3.6}}
 \put(89.5,10){\line(-1,-1){3.6}}
 \put(89.5,10){\circle*1}
 \put(103.5,10){\line(1,1){3.6}}
 \put(103.5,10){\line(1,-1){3.6}}
 \put(103.5,10){\circle*1}
 \put(112,9){\mbox + 3} 
 \put(130,12){\circle{10}} 
 \put(130,20){\oval(16,16)[b]}
 \put(125.25,13.2){\circle*1}
 \put(134.75,13.2){\circle*1}
 \put(130,7){\line(-1,-1)5}
 \put(130,7){\line(1,-1)5}
 \put(130,7){\circle*1}
\end{picture}
\caption{The vertex function of the $\phi^4$ theory in the two-loop
 approximation.} \label{fig1}
\end{figure}

Consider, for pedagogical purposes, the vertex function in the two-loop
approximation (see fig.\ref{fig1}).

We demonstrate the infrared peculiarities of the extension of
differential renormalization by the calculation of the
renormalization-group $\beta$ function. It coincides with the anomalous
dimension of four-point vertex up to the two-loop propagator
contribution ignored hereafter.

\subsection{Non-exceptional momenta} \indent

To simplify the calculation, we reduce the diagrams to the propagator
type by nullifying some external momenta. One can do this in two
different ways. The first way is to take non-exceptional external
momenta without creating any infrared divergencies. Then, according to
eqs.(\ref{31})--(\ref{311}), we have [Hereafter, we ignore obvious
factors of $\pi^2$ coming from the integrals, together with
$(2\pi)^{-4}$ from the loops, so that the contribution of each diagram
should be divided by $\left(16\pi^2\right)^n$, where $n$ is the number
of loops.]
\begin{eqnarray*} p~ \to
 \begin{picture}(16,9)(-8,-1) 
  \put(0,4.2){\oval(14.1,14.1)[b]}
  \put(0,-4.2){\oval(14.1,14.1)[t]}
  \put(-5.6,0){\circle*1}
  \put(5.6,0){\circle*1}
  \put(-5.9,-4){$\scriptstyle 0$}
  \put(4,-4){$x$}
 \end{picture}
 &=& \int \d^4 x \FR {\e^{\i p x}} {\left(x^2\right)^2}
 ~=~ - \ln (p^2/\mu^2) , \\
p~ \to
 \begin{picture}(22,9)(-11,-1) 
  \put(-3.5,0){\circle 7}
  \put(3.5,0){\circle 7}
  \put(-7,0){\line(-1,-1){3.6}}
  \put(-7,0){\line(-1,1){3.6}}
  \put(7,0){\line(1,1){3.6}}
  \put(7,0){\line(1,-1){3.6}}
  \put(0,0){\circle*1}
  \put(-7,0){\circle*1}
  \put(7,0){\circle*1}
  \put(-8,-5){$\scriptstyle 0$}
  \put(-1,-5){$y$} \put(6.5,-5){$x$}
 \end{picture}
 &=& \int \FR {\d^4 x~ \d^4 y~ \e^{\i p x}}
  {\left( y^2 \right)^2 \left[ (y-x)^2 \right]^2} =
  \left[ \int \d^4 x \FR {\e^{\i p x}} {\left(x^2\right)^2}
  \right]^2 = \ln^2 (p^2/\mu^2)~ , \\
p~ \to
 \begin{picture}(16,6)(-8,-1) 
  \put(0,0){\circle{9.8}}
  \put(-4.9,-4.9){\oval(9.8,9.8)[rt]}
  \put(-4.9,0){\circle*1}
  \put(4.9,0){\circle*1}
  \put(0,-4.9){\circle*1}
  \put(-4.9,0){\line(-1,0)2}
  \put(4.9,0){\line(1,0)2}
  \put(-8,-4){$\scriptstyle 0$}
  \put(-1,-8){$y$}
  \put(5.5,-4){$x$}
 \end{picture}
 &=& \int
 \FR {\d^4 x~ \d^4 y~ \e^{\i p x}} {x^2 (y-x)^2 \left(y^2\right)^2} \\*
&=& \int \FR {\d^4 x~ \e^{\i p x}} {x^2}
 \int \d^4 y \FR {1} {(y-x)^2} \FR {-\Box_y} {4}
 \left[ \FR {\ln (y^2\Mu^2)} {y^2}
 \right] \\*
&=& \int \d^4 x \FR {\e^{\i p x}} {x^2} \FR {\ln (x^2\Mu^2)} {x^2}
 = \fr 1 2 \ln^2 (p^2/\mu^2) - \ln (p^2/\mu^2) + 1~ .
\end{eqnarray*}
Summing up all these contributions, we get the invariant charge as
follows:
\begin{equation} \overline{h}_{\rm diff} = h
 + \fr 3 2 h^2 \ln (p^2/\mu^2) + h^3
 \left[ \fr 9 4 \ln^2 (p^2/\mu^2) - 3 \ln (p^2/\mu^2) + 3
 \right] . \label{313}
\end{equation}
The $\beta$ function is defined by
\begin{equation} \beta (h) = \mu^2 \FR {\d h} {\d\mu^2}, \label{314}
\end{equation}
while the invariant charge $\overline{h}$ is $\mu$-independent.
Differentiating eq.(\ref{313}) with respect to $\mu^2$, we get
\begin{equation} \beta (h) = \fr 3 2 h^2 - 3 h^3.  \label{315}
\end{equation}

Compare this calculation with the one that uses dimensional
regularization and the $\overline{\rm MS}$ scheme. One has
\begin{eqnarray*}
 \begin{picture}(16,9)(-8,-1) 
  \put(0,4.2){\oval(14.1,14.1)[b]}
  \put(0,-4.2){\oval(14.1,14.1)[t]}
  \put(-5.6,0){\circle*1}
  \put(5.6,0){\circle*1}
 \end{picture}
 &=& \int
 \FR {\d^{4-2\varepsilon}x~\e^{\i p x}}
  {\left(x^2\right)^{2-2\varepsilon}}
 = \FR 1 {\varepsilon~(1-2\varepsilon)}
 \left( \FR {\mu^2} {p^2}
 \right) ^{\displaystyle\varepsilon} , \\*
&& KR' = 1/\varepsilon~ , \s1 R = - \ln (p^2/\mu^2) + 2~ ; \\
 \begin{picture}(22,9)(-11,-1) 
  \put(-3.5,0){\circle 7}
  \put(3.5,0){\circle 7}
  \put(-7,0){\line(-1,-1){3.6}}
  \put(-7,0){\line(-1,1){3.6}}
  \put(7,0){\line(1,1){3.6}}
  \put(7,0){\line(1,-1){3.6}}
  \put(0,0){\circle*1}
  \put(-7,0){\circle*1}
  \put(7,0){\circle*1}
 \end{picture}
 &=&  \int
 \FR {\d^{4-2\varepsilon}x~\d^{4-2\varepsilon}y~\e^{\i p x}}
  {\left(y^2\right)^{2-2\varepsilon}
   {\left(y-x\right)^2}^{(2-2\varepsilon)}} =
 \left[
  \FR{\left(\mu^2/p^2\right)^\varepsilon} {\varepsilon~(1-2\varepsilon)}
 \right]^2 , \\*
&& R' =
 \left[ \FR {1} {\varepsilon~ (1-2\varepsilon)}
  \left( \FR {\mu^2} {p^2}
  \right) ^{\displaystyle\varepsilon}
 \right]^2 - \FR {2} {\varepsilon}
 \left[ \FR {1} {\varepsilon~ (1-2\varepsilon)}
  \left( \FR {\mu^2} {p^2}
  \right) ^{\displaystyle\varepsilon}
 \right] ,\\*
&& KR' = - 1/\varepsilon^2 , \s1
 R = \ln^2 (p^2/\mu^2) - 4 \ln (p^2/\mu^2) + 4~ ;
\end{eqnarray*}
\pagebreak[1]
$$
 \begin{picture}(16,6)(-8,-1) 
  \put(0,0){\circle{9.8}}
  \put(-4.9,-4.9){\oval(9.8,9.8)[rt]}
  \put(-4.9,0){\circle*1}
  \put(4.9,0){\circle*1}
  \put(0,-4.9){\circle*1}
  \put(-4.9,0){\line(-1,0)2}
  \put(4.9,0){\line(1,0)2}
 \end{picture}
 = \int
 \FR {\d^{4-2\varepsilon}x~\d^{4-2\varepsilon}y~\e^{\i p x}}
  {(y^2)^{2-2\varepsilon} \left[(y-x)^2\right]^{1-\varepsilon}
   \left(x^2\right)^{1-\varepsilon}}
 = \int
 \FR {\d^{4-2\varepsilon}x~\e^{\i p x}} {(x^2)^{1-\varepsilon}}
 \FR {1} {\varepsilon~(1-2\varepsilon)\left(x^2\right)^{1-2\varepsilon}}
 , $$
\begin{eqnarray*} R' &=& \int
 \FR {\d^{4-2\varepsilon}x~\e^{\i p x}}
  {\left(x^2\right)^{1-\varepsilon}}
 \left[
  \FR 1
   {\varepsilon~(1-2\varepsilon) \left(x^2\right)^{1-2\varepsilon}}
  -\FR {1} {\varepsilon \left(x^2\right)^{1-\varepsilon}}
 \right] \\*
&=&
 \FR {\left(\mu^2/p^2\right)^{2\varepsilon}}
  {2\varepsilon^2~(1-2\varepsilon)~(1-3\varepsilon)}
 -
 \FR {\left(\mu^2/p^2\right)^{2\varepsilon}}
  {\varepsilon^2~(1-2\varepsilon)} , \\*
KR' &=& - \FR 1 {2\varepsilon^2} + \FR 1 {2\varepsilon}, \s1
 R = \fr 1 2 \ln^2 (p^2/\mu^2) -3\ln (p^2/\mu^2) +\fr{11}2 .
\end{eqnarray*}
Thus, for the bare and invariant charges, we have, respectively,
\begin{equation}
 h_{_{\rm B}} = \left(\mu^2\right)^\varepsilon
 \left[ h + \fr 3 2 \left( h^2 - h^3 \right) /\varepsilon
  + \fr 9 4 h^3/\varepsilon^2
 \right] , \label{316}
\end{equation}
and
\begin{equation}
 \overline{h}_{_{\overline{\rm MS}}} = h + \fr 3 2 h^2
 \left[ \ln (p^2/\mu^2) - 2
 \right] + h^3
 \left[ \fr 9 4 \ln^2 (p^2/\mu^2) - 12~\ln (p^2/\mu^2) + \fr{39}2
 \right] . \label{317}
\end{equation}
We can now calculate the $\beta$ function, differentiating either
eq.(\ref{316}) or eq.(\ref{317}) with respect to $\mu^2$ and taking
into account eq.(\ref{314}). Both ways lead to eq.(\ref{315}) for the
$\beta$ function.

\subsection{Exceptional momenta in differential renormalization} \indent

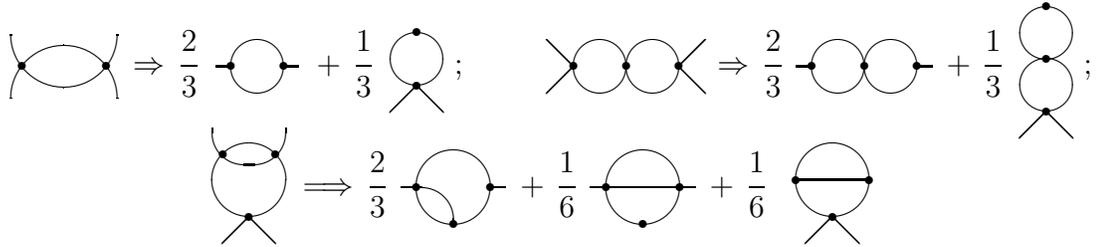
\begin{figure}[htbp] $$
 \begin{picture}(16,9)(-8,-1) 
  \put(0,4.2){\oval(14.1,14.1)[b]}
  \put(0,-4.2){\oval(14.1,14.1)[t]}
  \put(-5.6,0){\circle*1}
  \put(5.6,0){\circle*1}
 \end{picture}
 \Rightarrow \FR 2 3
 \begin{picture}(14,9)(-7,-1) 
  \put(0,0){\circle 7}
  \put(-3.5,0){\line(-1,0)2}
  \put(3.5,0){\line(1,0)2}
  \put(-3.5,0){\circle*1}
  \put(3.5,0){\circle*1}
 \end{picture}
 + \FR 1 3
 \begin{picture}(10,6)(-5,-2) 
  \put(0,0){\circle 7}
  \put(0,3.5){\circle*1}
  \put(0,-3.5){\circle*1}
  \put(0,-3.5){\line(-1,-1){3.6}}
  \put(0,-3.5){\line(1,-1){3.6}}
 \end{picture}
 ; \s1
 \begin{picture}(22,9)(-11,-1) 
  \put(-3.5,0){\circle 7}
  \put(3.5,0){\circle 7}
  \put(-7,0){\line(-1,-1){3.6}}
  \put(-7,0){\line(-1,1){3.6}}
  \put(7,0){\line(1,1){3.6}}
  \put(7,0){\line(1,-1){3.6}}
  \put(0,0){\circle*1}
  \put(-7,0){\circle*1}
  \put(7,0){\circle*1}
 \end{picture}
 \Rightarrow \FR 2 3
 \begin{picture}(20,9)(-10,-1) 
  \put(-3.5,0){\circle 7}
  \put(3.5,0){\circle 7}
  \put(-7,0){\line(-1,0)2}
  \put(7,0){\line(1,0)2}
  \put(0,0){\circle*1}
  \put(-7,0){\circle*1}
  \put(7,0){\circle*1}
 \end{picture}
 + \FR 1 3
 \begin{picture}(10,10)(-5,-2) 
  \put(0,3.5){\circle 7}
  \put(0,-3.5){\circle 7}
  \put(0,7){\circle*1}
  \put(0,0){\circle*1}
  \put(0,-7){\circle*1}
  \put(0,-7){\line(-1,-1){3.6}}
  \put(0,-7){\line(1,-1){3.6}}
 \end{picture}
 ; $$
$$
 \begin{picture}(12,7)(-6,-2) 
  \put(0,0){\circle{9.8}}
  \put(0,6.9){\oval(9.8,9.8)[b]}
  \put(-3.5,3.45){\circle*1}
  \put(3.5,3.45){\circle*1}
  \put(0,-4.9){\circle*1}
  \put(0,-4.9){\line(-1,-1){3.6}}
  \put(0,-4.9){\line(1,-1){3.6}}
 \end{picture}
 \Longrightarrow \FR 2 3
 \begin{picture}(16,6)(-8,-1) 
  \put(0,0){\circle{9.8}}
  \put(-4.9,-4.9){\oval(9.8,9.8)[rt]}
  \put(-4.9,0){\circle*1}
  \put(4.9,0){\circle*1}
  \put(0,-4.9){\circle*1}
  \put(-4.9,0){\line(-1,0)2}
  \put(4.9,0){\line(1,0)2}
 \end{picture}
 + \FR 1 6
 \begin{picture}(16,6)(-8,-1) 
  \put(0,0){\circle{9.8}}
  \put(-7,0){\line(1,0){14}}
  \put(-4.9,0){\circle*1}
  \put(4.9,0){\circle*1}
  \put(0,-4.9){\circle*1}
 \end{picture}
 + \FR 1 6
 \begin{picture}(16,7)(-8,-2) 
  \put(0,0){\circle{9.8}}
  \put(-4.9,0){\line(1,0){9.8}}
  \put(-4.9,0){\circle*1}
  \put(4.9,0){\circle*1}
  \put(0,-4.9){\circle*1}
  \put(0,-4.9){\line(-1,-1){3.6}}
  \put(0,-4.9){\line(1,-1){3.6}}
 \end{picture} $$
 \caption{Two-loop vertex diagrams in case of exceptional momenta.}
 \label{fig2}
\end{figure}

The second possibility is to take some exceptional momenta. In our case,
setting two of them equal to zero, we get the following decomposition
(see fig.\ref{fig2}).

Now, we ought to consistently remove infrared divergencies by defining
the artificial tadpole graphs which are both ultraviolet- and
infrared-divergent. We can use the invariance of the regularization, to
get a definition,
\begin{equation}
 \begin{picture}(10,6)(-5,-2) 
  \put(0,0){\circle 7}
  \put(0,3.5){\circle*1}
  \put(0,-3.5){\circle*1}
  \put(0,-3.5){\line(-1,-1){3.6}}
  \put(0,-3.5){\line(1,-1){3.6}}
 \end{picture}
 = \int \FR {\d^4 x} {\left(x^2\right)^2}
 ~ \stackrel {\strut\displaystyle\rm def} = ~
 \int \FR {\d^4 x} {\left(x^2\right)^2} \FR {x^2-2xy+y^2} {(x-y)^2}
 ~=~ -\ln (\nu^2/\mu^2) - 2~ , \label{318}
\end{equation}
where we have evaluated the integrals, according to
eqs.(\ref{31})--(\ref{312}).

Consider now the two-loop diagrams. We have
\begin{eqnarray*}
 \begin{picture}(10,10)(-5,-2) 
  \put(0,3.5){\circle 7}
  \put(0,-3.5){\circle 7}
  \put(0,7){\circle*1}
  \put(0,0){\circle*1}
  \put(0,-7){\circle*1}
  \put(0,-7){\line(-1,-1){3.6}}
  \put(0,-7){\line(1,-1){3.6}}
 \end{picture}
 &=& \int \FR {\d^4 x} {\left(x^2\right)^2}
 \int \FR {\d^4 y} {\left[ (x-y)^2 \right]^2} =
 \left[ \ln (\nu^2/\mu^2) + 2
 \right]^2 , \\[3mm]
 \begin{picture}(16,6)(-8,-1) 
  \put(0,0){\circle{9.8}}
  \put(-7,0){\line(1,0){14}}
  \put(-4.9,0){\circle*1}
  \put(4.9,0){\circle*1}
  \put(0,-4.9){\circle*1}
 \end{picture}
 &=& \int \d^4 x \FR {\e^{\i p x}} {\left(x^2\right)^2}
 \int \FR {\d^4 y} {y^2~(y-x)^2} =
 \int \d^4 x \FR {\e^{\i p x}} {\left(x^2\right)^2}
 \int \FR {\d^4 q} {\left(q^2\right)^2} \e^{-\i q x} \\*
&=& -\int \d^4 x \FR {\e^{\i p x}} {\left(x^2\right)^2} \ln (x^2\Nu^2)
 = -\int \d^4 x \FR {\e^{\i p x}} {\left(x^2\right)^2}
 \left[ \ln (x^2\Mu^2) + \ln (\nu^2/\mu^2)
 \right] \\*
&=& -\fr 1 2 \ln^2 (p^2/\mu^2) + \ln (p^2/\mu^2)~ \ln (\nu^2/\mu^2)
 + \ln (p^2/\mu^2) - 1~ .
\end{eqnarray*}
As for the last diagram of fig.\ref{fig2}, it can be calculated in two
different ways: first, integrating over $y$ in a usual way, and then
considering the $x$ tadpole-type integral; or conversely. Proceeding in
both ways, we get, respectively,
\pagebreak[1]
$$
 \begin{picture}(16,7)(-8,-2) 
  \put(0,0){\circle{9.8}}
  \put(-4.9,0){\line(1,0){9.8}}
  \put(-4.9,0){\circle*1}
  \put(4.9,0){\circle*1}
  \put(0,-4.9){\circle*1}
  \put(0,-4.9){\line(-1,-1){3.6}}
  \put(0,-4.9){\line(1,-1){3.6}}
  \put(-8,0){$\scriptstyle 0$}
  \put(6.5,0){$y$}
  \put(-1,-9){$x$}
 \end{picture}
 ~=~ \int \FR {\d^4 x~ \d^4 y} {\left(y^2\right)^2~ (y-x)^2~ x^2}
 = \int \d^4 x \FR {\ln (x^2\Mu^2)} {\left(x^2\right)^2} \s3 $$
\begin{equation} = \int \d^4 x \FR {\ln (x^2\Mu^2)} {\left(x^2\right)^2}
 \FR {x^2-2xz+z^2} {(x-z)^2}
 = \fr 1 2 \ln^2 (\nu^2/\mu^2) + \ln (\nu^2/\mu^2)~ , \label{324}
\end{equation}
\pagebreak[1]
$$
 \begin{picture}(16,7)(-8,-2) 
  \put(0,0){\circle{9.8}}
  \put(-4.9,0){\line(1,0){9.8}}
  \put(-4.9,0){\circle*1}
  \put(4.9,0){\circle*1}
  \put(0,-4.9){\circle*1}
  \put(0,-4.9){\line(-1,-1){3.6}}
  \put(0,-4.9){\line(1,-1){3.6}}
 \end{picture}
 ~=~ \int \FR {\d^4 x~\d^4 y} {\left(y^2\right)^2~(y-x)^2~x^2}
 = - \int \d^4 y \FR {\ln (y^2\Nu^2)} {\left(y^2\right)^2} \s3 $$
$$ = - \int \d^4 y
 \FR {\ln(y^2\Mu^2)+\ln(\nu^2/\mu^2)} {\left(y^2\right)^2}
 = \fr 1 2 \ln^2 (\nu^2/\mu^2) + \ln (\nu^2/\mu^2) . \eqnum{\ref{324}'}
 $$
Coincidence of these two expressions gives us, thus, a self-consistency
check of the definition of the artificial tadpole graph (\ref{318}).

Combining these two-loop contributions with those already found for
non-ex\-cep\-tion\-al momenta, we finally get the invariant charge
\begin{eqnarray} && \overline{h}^{\rm(ex)}_{\rm diff} = h + h^2
 \left[ \ln (p^2/\mu^2) + \fr 1 2 \ln (\nu^2/\mu^2) + 1
 \right] + h^3
 \left[ \fr 5 4 \ln^2 (p^2/\mu^2)
 \right. \label{319} \\*
&& ~~
 \left. + \fr 1 2 \ln (p^2/\mu^2)~ \ln (\nu^2/\mu^2)
  + \fr 1 2 \ln^2 (\nu^2/\mu^2) - \fr 3 2 \ln (p^2/\mu^2)
  + \fr 3 2 \ln (\nu^2/\mu^2) + \fr 5 2
 \right] .~~~~ \nonumber
\end{eqnarray}

We see that eq.(\ref{319}) differs from eq.(\ref{313}). However,
differentiating it with respect to $\mu^2$, we obtain the $\beta$
function in agreement with eq.(\ref{315}).

\subsection{Exceptional momenta in dimensional renormalization} \indent

To get a better understanding of the situation with exceptional momenta,
we again compare the extended differential renormalization with
dimensional regularization in the $\overline{\rm MS}$ scheme. In the
latter case, we use the infrared $\tilde{R}$ operation to remove
infrared divergencies:
\begin{eqnarray*}
 \begin{picture}(10,6)(-5,-2) 
  \put(0,0){\circle 7}
  \put(0,3.5){\circle*1}
  \put(0,-3.5){\circle*1}
  \put(0,-3.5){\line(-1,-1){3.6}}
  \put(0,-3.5){\line(1,-1){3.6}}
 \end{picture}
 &=& \int \FR {\d^{4-2\varepsilon}p} {\left(p^2\right)^2} , \s1
 \tilde{R} =
 \left(\mu^2\right)^\varepsilon \int \d^{4-2\varepsilon}p
 \left[ \FR 1 {\left(p^2\right)^2} +
  \FR {\delta_{4-2\varepsilon}(p)}
   {\varepsilon \left(\nu^2\right)^\varepsilon}
 \right] = \FR 1 \varepsilon
 \left( \FR {\mu^2} {\nu^2}
 \right) ^{\displaystyle\varepsilon} , \\*
&& KR'\tilde{R} = 1/\varepsilon, \s1 R^* = \ln (\mu^2/\nu^2) ; \\
 \begin{picture}(10,10)(-5,-2) 
  \put(0,3.5){\circle 7}
  \put(0,-3.5){\circle 7}
  \put(0,7){\circle*1}
  \put(0,0){\circle*1}
  \put(0,-7){\circle*1}
  \put(0,-7){\line(-1,-1){3.6}}
  \put(0,-7){\line(1,-1){3.6}}
 \end{picture}
 &=& \int \FR {\d^{4-2\varepsilon}p} {\left(p^2\right)^2}
 \int \FR {\d^{4-2\varepsilon}q} {\left(q^2\right)^2} , \s2 \tilde{R} =
 \left[ \FR 1 \varepsilon
  \left( \FR {\mu^2} {\nu^2}
  \right) ^{\displaystyle\varepsilon}
 \right]^2 , \\*
&& R'\tilde{R} = \FR 1 {\varepsilon^2}
 \left( \FR {\mu^2} {\nu^2}
 \right)^{2\displaystyle\varepsilon} - \FR 2 {\varepsilon^2}
 \left( \FR {\mu^2} {\nu^2}
 \right) ^{\displaystyle\varepsilon}, ~~
 KR'\tilde{R} =  - 1/\varepsilon^2, ~~
 R^* = \ln^2 (\mu^2/\nu^2)~ ;
\end{eqnarray*}
\pagebreak[1]
$$ \tilde{R}R'
 \begin{picture}(16,6)(-8,-1) 
  \put(0,0){\circle{9.8}}
  \put(-7,0){\line(1,0){14}}
  \put(-4.9,0){\circle*1}
  \put(4.9,0){\circle*1}
  \put(0,-4.9){\circle*1}
 \end{picture}
 = \left(\mu^2\right)^\varepsilon \int \d^{4-2\varepsilon}p
 \left[ \FR 1 {\left(p^2\right)^2} + \FR 1 \varepsilon
  \FR {\delta_{4-2\varepsilon}(p)} {\left(\nu^2\right)^\varepsilon}
 \right]
 \left[ \FR 1 {\varepsilon~ (1-2\varepsilon)}
  \FR {\left(\mu^2\right)^\varepsilon}
   {\left[(p-q)^2\right]^\varepsilon}
  - \FR 1 \varepsilon
 \right] $$
$$ = - \FR {\left(\mu^2/p^2\right)^{2\varepsilon}}
 {2\varepsilon^2~(1-3\varepsilon)} +
 \FR {\left(\mu^2/p^2\right)^\varepsilon
   \left(\mu^2/\nu^2\right)^\varepsilon}
  {\varepsilon^2~(1-2\varepsilon)}
 - \FR {\left(\mu^2/\nu^2\right)^{2\varepsilon}} {\varepsilon^2} , $$
$$ KR'\tilde{R} = - \FR {1} {2\varepsilon^2} +
 \FR {1} {2\varepsilon} ,~~~  R^* = - \fr 1 2 \ln^2 \FR {p^2} {\mu^2}
 + \ln \FR {p^2} {\mu^2} \ln \FR {\nu^2} {\mu^2}
 + \ln \FR {p^2} {\mu^2} - 2\ln \FR {\nu^2} {\mu^2} - \fr 1 2 ; $$
\begin{eqnarray*} R'
 \begin{picture}(16,7)(-8,-2) 
  \put(0,0){\circle{9.8}}
  \put(-4.9,0){\line(1,0){9.8}}
  \put(-4.9,0){\circle*1}
  \put(4.9,0){\circle*1}
  \put(0,-4.9){\circle*1}
  \put(0,-4.9){\line(-1,-1){3.6}}
  \put(0,-4.9){\line(1,-1){3.6}}
 \end{picture}
 &=& \left(\mu^2\right)^\varepsilon \int \d^{4-2\varepsilon}p
 \left[ \FR 1 {\varepsilon~ (1-2\varepsilon)}
  \FR {(\mu^2)^\varepsilon} {\left(p^2\right)^{2+\varepsilon}}
  - \FR 1 \varepsilon \FR {1} {\left(p^2\right)^2}
 \right] , \\*
\tilde{R}R' &=& \left(\mu^2\right)^\varepsilon
 \int \d^{4-2\varepsilon}p
 \left[ \FR 1 {\varepsilon~ (1-2\varepsilon)}
  \FR {\left(\mu^2\right)^\varepsilon}
   {\left(p^2\right)^{2+\varepsilon}}
  + \FR {1+\varepsilon} {2\varepsilon^2}
  \FR {\left(\mu^2\right)^\varepsilon}
   {\left(\nu^2\right)^{2\varepsilon}}
  \delta_{4-2\varepsilon}(p)
 \right. \\*
&&
 \left. - \FR {1} {\varepsilon} \FR {1} {\left(p^2\right)^2}
  - \FR {1} {\varepsilon^2}
  \FR {\delta_{4-2\varepsilon}(p)} {\left(\nu^2\right)^\varepsilon}
 \right] ~=~ \FR {1+\varepsilon} {2\varepsilon^2}
 \left( \FR {\mu^2} {\nu^2}
 \right)^{2\displaystyle\varepsilon}
 - \FR {1} {\varepsilon^2}
 \left( \FR {\mu^2} {\nu^2}
 \right) ^{\displaystyle\varepsilon}, \\*
KR'\tilde{R} &=& - \FR {1} {2\varepsilon^2} +
 \FR {1} {2\varepsilon} , \s2
 R^* =  \fr 1 2 \ln^2 (\mu^2/\nu^2) + \ln (\mu^2/\nu^2)~ .
\end{eqnarray*}

Combining everything together, we get for the bare charge precisely
eq.(\ref{316}), and for the invariant charge
\begin{eqnarray} &&
 \overline{h}^{\rm(ex)}_{_{\overline{\rm MS}}} = h + h^2
 \left[ \ln (p^2/\mu^2) + \fr 1 2 \ln (\nu^2/\mu^2) - 2
 \right] + h^3
 \left[ \fr 5 4 \ln^2 (p^2/\mu^2)
 \right. \label{321} \\*
&& ~~
 \left. + \fr 1 2 \ln (p^2/\mu^2)~ \ln (\nu^2/\mu^2)
  + \fr 1 2 \ln^2 (\nu^2/\mu^2) - \fr{15}2 \ln (p^2/\mu^2)
  - \fr 3 2 \ln (\nu^2/\mu^2) + \fr{51}4
 \right] .~~~~ \nonumber
\end{eqnarray}

Note that eq.(\ref{321}) does not coincide with eq.(\ref{319}). However,
if we differentiate eq.(\ref{321}) with respect to $\mu^2$, we reproduce
the correct value of the $\beta$ function eq.(\ref{315}).

Hence, in dimensional regularization, when one uses $R^*$ operation, the
$\beta$ function calculated both from infinite and finite parts is
independent of external momenta in the $\overline{\rm MS}$ scheme. This
is also true for the finite expressions obtained with the aid of the
extended differential renormalization, which give us the correct value
of the $\beta$ function despite the fact that the finite corrections to
the invariant charge eqs.(\ref{313}) and (\ref{319}) are different from
those in the $\overline{\rm MS}$ scheme, eqs.(\ref{317}) and
(\ref{321}). This difference is, however, trivial and reflects just the
renormalization scheme arbitrariness. Moreover, to two loops, the finite
part of each diagram in the extended differential renormalization can be
obtained, up to a constant, from the corresponding $\overline{\rm MS}$
renormalized expression by a shift: $\ln \mu^2$$\to$$\ln \mu^2$$-$2.
This statement happens to be independent of the presence or absence of
infrared divergencies.

\subsection{Higher-order ambiguities} \indent

This pleasant picture is, however, spoiled in higher orders. To see
this, let us consider the following four-loop artificial tadpole graph:
$$
 \begin{picture}(18,9)(-9,-2) 
  \put(0,0){\circle{14.1}}
  \put(0,5){\oval(10,4.2)[b]}
  \put(0,-7){\line(-1,-1){3.6}}
  \put(0,-7){\line(1,-1){3.6}}
  \put(0,-7){\circle*1}
  \put(-5,5){\circle*1}
  \put(5,5){\circle*1}
  \put(-6.9,1.5){\circle*1}
  \put(-4.8,-5.2){\circle*1}
  \put(-6,-1.9){\circle 7}
  \put(-10,1.5){$t$}
  \put(-6.5,-8){$z$}
  \put(-7.5,5.5){$\scriptstyle 0$}
  \put(6.5,5.5){$y$}
  \put(-1,-11){$x$}
 \end{picture} \vspace{3mm}
 = \int
 \FR {\d^4 x~ \d^4 y ~\d^4 z~ \d^4 t}
  {\left(y^2\right)^2 (y-x)^2~ (x-z)^2 \left[(z-t)^2\right]^3 t^2} . $$
To evaluate the internal propagator-type subgraph, we need to add a new
equation to the set of basic formulae, eqs.(\ref{31})--(\ref{312}),
\begin{equation} \FR 1 {\left(x^2\right)^3} = - \FR 1{32} \Box_x^2
 \FR {\ln (x^2\Mu^2)} {x^2} + \FR 3{16} \pi^2~ \Box_x~ \delta_4 (x)~,
 \label{322}
\end{equation}
where the coefficient of the last term is strictly fixed by the
requirement of consistency with the rest of the formulae after
multiplying both sides by $x^2$. Performing the internal integration
with the aid of eq.(\ref{322}), we reduce the diagram to the following
integral,
\begin{equation} I = -\fr 1 2 \int
 \FR {\d^4 x~ \d^4 y~ \left[ \ln (x^2\Mu^2)+\fr 3 2 \right]}
  {\left(y^2\right)^2 (y-x)^2~ x^2} , \label{325}
\end{equation}
which, again, can be integrated first over $y$ and then over $x$, or in
the opposite order. We have, respectively,

\begin{eqnarray} I &=& -\fr 1 2 \int \d^4 x
 \FR {\ln(x^2\Mu^2)\left[\ln(x^2\Mu^2)+\fr{3}2\right]}
  {\left(x^2\right)^2}
 \nonumber\\*
&=& -\fr 1 2 \int \d^4 x
 \FR {\ln(x^2\Mu^2)\left[\ln(x^2\Mu^2)+\fr{3}2\right]}
  {\left(x^2\right)^2}
 \FR {x^2-2xz+z^2} {(x-z)^2} \nonumber\\*
&=& \fr 1 6 \ln^3 (\nu^2/\mu^2) + \fr 1 8 \ln^2 (\nu^2/\mu^2)
 - \fr 3 4 \ln (\nu^2/\mu^2) + \fr 2 3 \zeta (3)~ , \label{323}
\end{eqnarray}
\pagebreak[1]
\begin{eqnarray*} I &=& \fr 1 2 \int \FR {\d^4 y} {\left(y^2\right)^2}
 \int \FR {\d^4 q} {q^2} \e^{-\i q y} \FR {\ln(q^2/\mu^2)-\fr{3}2} {q^2}
 \\*
&=& \fr 1 2 \int \FR {\d^4 y} {\left(y^2\right)^2}
 \Bigl\{ \fr 1 2 \ln^2 (y^2\Nu^2) +
  \left[ \fr 1 2 - \ln (\nu^2/\mu^2)
  \right] \ln (y^2\Nu^2) + 1
 \Bigr\}
\end{eqnarray*}
$$ = \fr 1 6 \ln^3 (\nu^2/\mu^2) + \fr 1 8 \ln^2 (\nu^2/\mu^2)
 - \fr 3 4 \ln (\nu^2/\mu^2) - \fr 1 3 \zeta (3) - 1~ .
 \eqnum{\ref{323}'} $$
In contrast to the previous case of eqs.(\ref{324}), the two expressions
(\ref{323}) and (\ref{323}$'$) are different. It turns out to be a
general situation, which simply has not yet manifested itself at the
two-loop level but becomes evident at the fourth loop [In fact,
integrals like eq.(\ref{325}) appear already at the three-loop level].
Thus, we see that, in $D$=4 either, the infrared extension of
differential renormalization does not provide us with a self-consistent
definition of tadpole graphs. On the other hand, abandoning the infrared
$\tilde{R}$ operation and just nullifying artificial tadpoles, we would
not be able to make use of infrared rearrangement at all.

Therefore, we conclude that the infrared extension of differential
renormalization cannot perform the infrared $\tilde{R}$ operation, like
the original scheme does the ultraviolet $R$ operation.

\section{Summary} \indent

We have examined the possibility of generalizing differential
regularization in an invariant fashion to theories with infrared
divergencies. Both in $D$=2 and $D$=4, the basic differential identities
of the method, written in co-ordinate and momentum space, lead to a
definite set of consistent formulae for divergent Fourier integrals
regularized by two scale parameters, $\mu$ and $\nu$, which remove
ultraviolet and infrared singularities, respectively.

The principle of invariance of the regularization allows us also to
derive definitions for the tadpole-type integrals (without external
momenta or co-ordinates), which intermix infrared and ultraviolet
divergencies, by splitting the integrals up into a sum of separate
infrared, ultraviolet, and regular items. However, the values for
tadpole graphs of a complicated structure in higher orders of
perturbation theory prove to be ambiguous, depending on the order of
evaluation of their subgraphs. This ambiguity cannot be fully eliminated
even if we try to somehow relate $\nu$ with $\mu$ (which makes low-order
tadpole diagrams to be zero, but nevertheless, does not lead to a unique
determination of higher-order graphs; moreover, any information about
ultraviolet logarithms is completely lost then).

In two-dimensional $\sigma$ models, where intrinsic infrared
divergencies at zero masses are present, the results of the extended
differential renormalization disagree with the finite minimally
renormalized results of dimensional regularization. The difference goes
beyond the range allowed by renormalization-scheme arbitrariness in
perturbation theory. For the $O(n)$ sigma model with a single coupling
constant, the two-loop coefficient of the $\beta$ function turns out to
be zero when we use the infrared extension of differential
renormalization. This contradicts the old well-established result which
should be scheme-independent.

As concerns the use of the infrared $\tilde{R}$ operation in dimensional
regularization, we have verified by direct two-loop calculations that in
renormalizable theories ($\vec{\bf n}$-field in $D$=2$-$2$\varepsilon$
and $\phi^4$ in $D$=4$-$2$\varepsilon$) it successfully recovers
ultraviolet logarithms in the finite parts of the diagrams with
exceptional momenta, so that the $\beta$ function, calculated through
the finite invariant charge, proves to be independent of any logarithms
(of momenta and renormalization scales). It coincides with the result
derived in the conventional way from the coefficients of the first
singularity in $\varepsilon$, which does not depend on the choice of the
momenta for logarithmically divergent diagrams.

The two-loop results of the infrared extension of differential
renormalization in the $\phi^4_{(D=4)}$ theory agree with dimensional
renormalization up to finite counterterms. However, higher-order
logarithmically divergent artificial tadpole diagrams (resulting from
infrared rearrangement) suffer from an irremovable ambiguity. Thus,
calculations with exceptional momenta cannot be performed consistently
above two loops.

Our final conclusion is that the program of constructing an invariant
generalization of differential regularization and renormalization, to
deal with infrared divergencies, has failed.

\end{document}